\documentclass[aip,jcp,reprint,superscriptaddress,amsmath,amssymb,footinbib,floatfix]{revtex4-2}

\usepackage{graphicx}
\usepackage{dcolumn}
\usepackage{bm}

\usepackage{lipsum}
\newcommand{\deltxt}[1]{}
\newcommand{\newtxt}[1]{#1}

\newcommand{\fref}[1]{Figure~\ref{#1}}

\newcommand{\eref}[1]{Equation~\ref{#1}}
\newcommand{\cref}[1]{Chapter~\ref{#1}}

\begin{document}

\preprint{APS/123-QED}

\title{Flow dichroism of DNA can be quantitatively predicted via coarse-grained molecular simulations}

\author{Isaac Pincus}
\altaffiliation{Present address: Department of Chemical Engineering, Massachusetts Institute of Technology, Cambridge, MA 02139, USA}
\affiliation{Department of Chemical and Biological Engineering, Monash University, Clayton, VIC 3800, Australia
}%

\author{Alison Rodger}
\affiliation{
Research School of Chemistry, Australian National University, ACT 2601, Australia
}%

\author{J. Ravi Prakash}
\email{To whom correspondence should be addressed. E-mail: ravi.jagadeeshan@monash.edu}
\affiliation{Department of Chemical and Biological Engineering, Monash University, Clayton, VIC 3800, Australia
}%

\date{\today}

\begin{abstract}
We demonstrate the use of multiscale polymer modelling to quantitatively predict DNA linear dichroism (LD) in shear flow. 
LD is the difference in absorption of light polarised along two perpendicular axes, and has long been applied to study biopolymer structure and drug-biopolymer interactions. 
As LD is orientation-dependent, the sample must be aligned in order to measure a signal.
Shear flow via a Couette cell can generate the required orientation, however it is challenging to separate the LD due to changes in polymer conformation from specific interactions, e.g. drug-biopolymer.
In this study, we have applied a combination of Brownian dynamics and equilibrium Monte Carlo simulations to accurately predict polymer alignment, and hence flow LD, at modest computational cost. 
As the optical and conformational contributions to the LD can be explicitly separated, our findings allow for enhanced quantitative interpretation of LD spectra through the use of an in-silico model to capture conformational changes.
Our model requires no fitting and only five input parameters, the DNA contour length, persistence length, optical factor, solvent quality, and relaxation time, all of which have been well characterized in prior literature. 
The method is sufficiently general to apply to a wide range of biopolymers beyond DNA, and our findings could help guide the search for new pharmaceutical drug targets via flow LD.
\end{abstract}

\maketitle


\section{Introduction}

In recent years, there has been growing interest in the development of high-throughput drug screening techniques to address the challenges posed by emerging antibiotic resistance and novel viral strains. 
One such technique is linear dichroism (LD), which has traditionally been used for studying macromolecular structures \cite{rodger2016linear, Rodger2009, norden2019linear}. 
LD has gained attention as a promising method for rapidly determining the extent of drug binding to biological structures of interest, including proteins, lipid membranes, or DNA/RNA, and to provide information about how drugs affect the structure of the biological target \cite{Broughton2016}.
Unlike several other techniques (e.g. crystallography, electron microscopy), LD can be employed on systems in solution under biologically relevant temperatures and conditions. 
This enables the investigation of interactions without the need for extensive sample preparation and disruption.
However, it often lacks quantitative accuracy due to the need to separate the LD signal caused by the orientation of a specific macromolecule from the signal resulting from its interaction with a target drug. 
For example, the orientation of intercalating dyes interacting with DNA can be determined through LD measurements, but only if the orientation of the DNA can be deconvoluted from the overall signal.

While the decoupling of orientation and interaction contributions to LD is a challenging problem to address experimentally \cite{McLachlan2013, rodger2016linear}, recent improvements to \textit{in silico} Brownian dynamics (BD) simulation techniques have shown that it is possible to predict the elongation and orientation of macromolecules in flow using coarse-grained models \cite{sasmal2017parameter, Prakash2019review}.
In this paper, we showcase the potential of BD techniques in conjunction with equilibrium Monte Carlo (MC) methods to accurately capture the experimentally measured LD signal of DNA solutions at various molecular weights and shear rates \cite{Kubista1993}. 
The versatility of this technique suggests its potential application to any linear macromolecule that can be described using a coarse-grained model based on equilibrium structural characteristics and deformation timescales, including DNA/RNA \newtxt{(both single-stranded and double-stranded)}, structural proteins, or filamentous bacteriophages.
\newtxt{
Although the current work seeks to reproduce prior experimental results for dilute DNA in a buffer solution, we note that more complicated interactions have been incorporated into Brownian dynamics simulations, for example coarse-grained receptor-ligand binding using Monte Carlo acceptance algorithms \cite{robe2024evanescent}, Ewald summation methods for hydrodynamics of large systems with multiple species \cite{fiore2019fast}, or active motor forces to drive cooperative alignment \cite{peruani2016active}.
}

\begin{figure}[t]
  \centering
  \includegraphics[width=8.5cm]{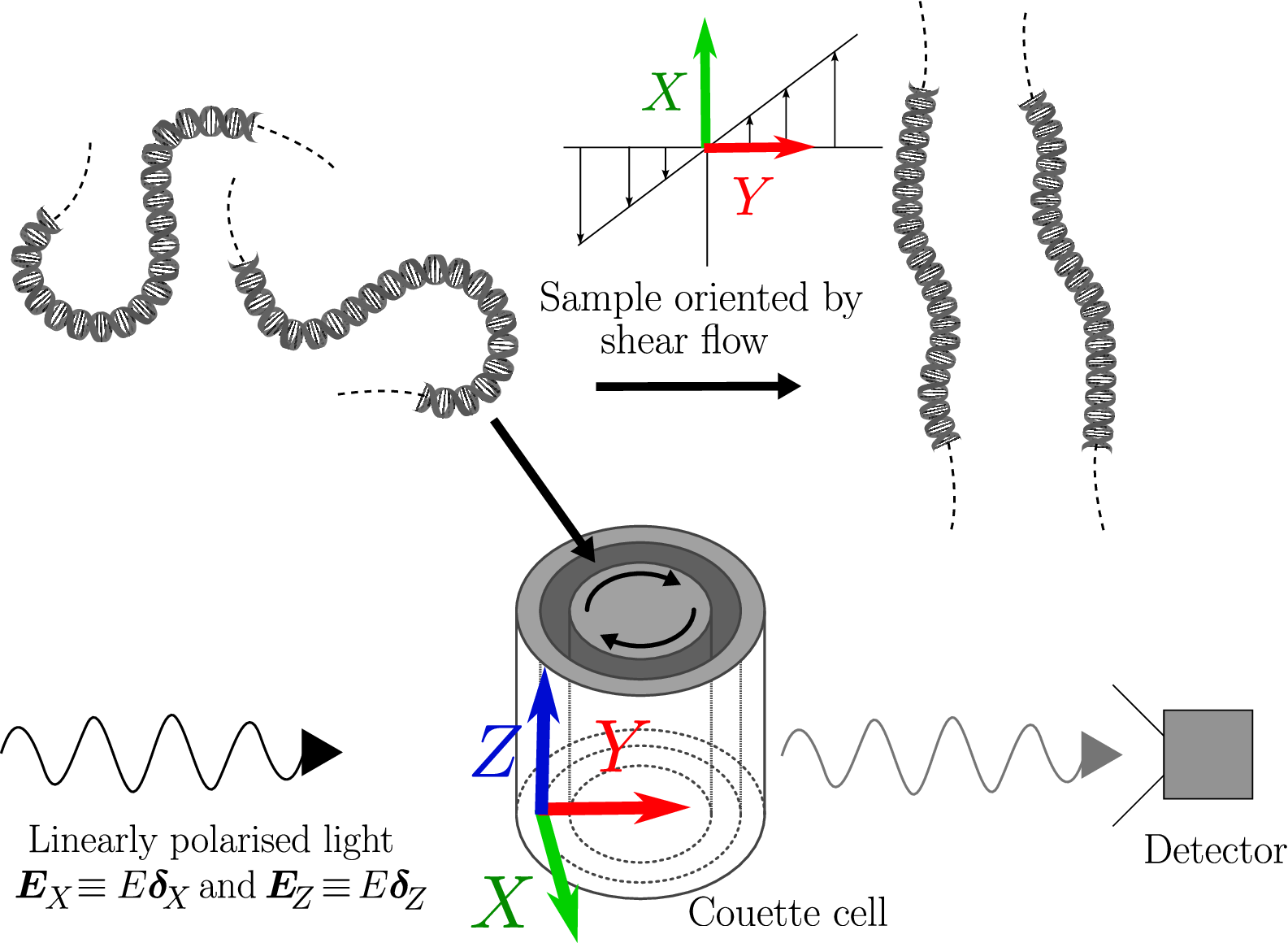}
  \caption[Couette Cell for Shear Flow LD]{Use of a Couette cell for orientation of a sample. Shear flow within the cell aligns the molecules, allowing a net LD signal to be measured. In this case, as in experiments by Simonson and Kubista \cite{Kubista1993}, light is propagated radially through the cell (the $Y$ direction) and LD is measured with light polarised axially ($Z$ direction) and perpendicular to the $Y-Z$ plane (the $X$ direction). Relative to the flow field between the cylinders, the light is propagated along the shear gradient direction (the laboratory axis $Y$), while LD is measured along the flow direction ($X$) and neutral direction ($Z$).}
  \label{Couette Cell LD}
\end{figure}

More precisely, LD is the polarisation-dependent absorption of light by an ensemble of oriented molecules, such that
\begin{equation}
\label{eq: LD parallel perp}
    \mathrm{LD}^\mathrm{r} = \frac{A_{\parallel} - A_{\bot}}{A_\mathrm{iso}}
\end{equation}
where $A_{\parallel}$ and $A_{\bot}$ represent the absorption of light polarised in perpendicular directions relative to some laboratory axis (the $\parallel$ direction). 
$A_\mathrm{iso}$ is the isotropic absorbance of the sample, and the $\mathrm{r}$ denotes `reduced' dichroism. 
Flow LD (\fref{Couette Cell LD}) is able, e.g., to probe reaction kinetics such as assembly of protein fibres \cite{MontgomeryPhD, Dafforn2004}, cleavage of DNA, or protein-membrane interactions \cite{rodger2016linear} with relatively small sample volumes (70 $\mu$L) \cite{marrington2004}. 
However, sample orientation under shear flow is both imperfect and configurationally complex due to the combination of rotational and elongational velocity components - thus, flow LD data interpretation is generally restricted to being qualitative or semi-quantitative, and attempts to calculate the orientation parameter have been limited to either small or highly rigid macromolecules \cite{McLachlan2013, wilson1978flow, OdegaardJensen1996}.
If this difficulty could be overcome, and sample orientations determined for a particular macromolecular sample at a particular shear rate, it would be possible to considerably improve the quantitative accuracy of LD spectroscopy analysis \cite{Rodger2009, rodger2016linear}.

Since a full analytical theory for the conformation of a flexible polymer chain in shear flow is not possible due to nonlinear coupling with the solvent, previous treatments have relied on approximations regarding the chain connectivity, perturbation due to shear flow, and physical effects such as solvent-polymer interactions and hydrodynamic forces on beads \cite{norden_kubista_kurucsev_1992, Larson2005review, Prakash2019review}.
Advances in modelling of dilute polymer solutions, most notably the development and refinement of simulation methods such as Brownian dynamics (BD), allow one to avoid many of the earlier approximations \cite{Ottinger1996}.
It is now possible to qualitatively recover much of the key behaviour of polymers in shear flows, and even obtain quantitative, parameter-free predictions in extensional flows \cite{Prakash2019review, Prabhakar2004SFG, Saadat2015SFG, Sunthar2005parameterfree}.

While LD is measurable for many important biopolymers, in the current work we focus our attention on DNA, as it is widely studied and we have the experimental data of Simonson and Kubista \cite{Kubista1993} with which to compare our simulation results.
We apply a recently-developed polymer model based upon the so-called FENE-Fraenkel spring \cite{pincus2020viscometric, Pincus2023} which can represent a DNA fragment of arbitrary length, from 300 base pairs (bp) all the way to 164 kilo-base pairs (kbp), by `coarse-graining' the underlying polymer as a chain of beads connected by springs.
Rather than setting a constant 10 base pairs per segment as in previous BD simulations of LD \cite{OdegaardJensen1996}, we use a spring force law which can represent anything from tens of base pairs to thousands of base pairs, so that one can simulate even long DNA fragments ($> 100$ kbp) with less than 100 springs, which allows for efficient simulation (since computational cost grows as $N^{2.25}$ \cite{Prabhakar2004Separation}).
Our challenge is to apply such simulations to experimental LD data, with the aforementioned separation of optical and conformational components of the LD signal.
The problem of determining the LD of a real polymer chain given a coarse-grained representation has been investigated in prior literature \cite{wilson1977dichroic, wilson1978flow, kuhn1942relationships, Yamakawa1971}.
However, as previously mentioned, these early treatments required significant approximations to obtain closed-form results which limit their applicability to only long, flexible polymer chains at low shear rates, or very short, rigid molecules.

By applying our polymer model and performing a combination of BD and MC simulations, we are able to quantitatively match the experimentally-measured LD of different length DNA fragments at different shear rates.
This requires not only BD modelling but also accounting for the average orientation of transition moments in the coarse-grained segments.
Improving on previous analytical \cite{norden_kubista_kurucsev_1992} and simulation-based \cite{OdegaardJensen1996} methods, our model works for arbitrary shear rates and DNA lengths.
Furthermore, when dividing the polymer chain into coarse-grained segments, we show that the overall LD signal can be described through the separable influence of segment orientation and segment extension.

\section{Methods}
\label{LD theory}

LD is an absorption spectroscopy and so arises from the coupling of the electric field vector, $\bm{E}$, of light with an electric transition dipole moment $\bm{\mu}$ of a molecule to cause transitions between molecular energy states.
$\bm{\mu}$ is an integral function of the electric dipole operator and the initial and final molecular wavefunctions for the transition \cite{fuller1995optical}.
The oscillator strength $A$ (or absorption magnitude) of a single dipole $\bm{\mu}$ with respect to an electric field $\bm{E}$ may be written:
\begin{equation}
\label{basic LD equation}
    A(\mu, E) = k |\bm{\mu} \cdot \bm{E}|^2 = k (\mu E \cos{\Omega})^2
\end{equation}
where $k$ is a constant, $\mu$ and $E$ are the magnitudes of the $\bm{\mu}$ and $\bm{E}$ vectors, while $\Omega$ is the angle between them \cite{Rodger2009, norden_kubista_kurucsev_1992}.
The signal in a real solution is an ensemble average over the many molecules which interact with the incident light electric field (in other words, an ensemble average over all $\bm{\mu}$ at constant $\bm{E}$). 

If one were able to determine the position and orientation of every single transition dipole moment $\bm{\mu}$ in some solution of DNA, it would be in principle straightforward to use \eref{eq: LD parallel perp} and \eref{basic LD equation} to calculate the $LD$ by averaging the absorption $A$ over all $\bm{\mu}$.
The challenge is to locate each $\bm{\mu}$, which is in general extremely complex given an ensemble of polymer chains in shear flow.
However, each $\bm{\mu}$ is not freely floating in space, but instead attached to the DNA double-helix, and so it is possible to decompose the overall LD signal into separate components.

\begin{figure*}[!ht]
  \centering
  \includegraphics[width=\textwidth]{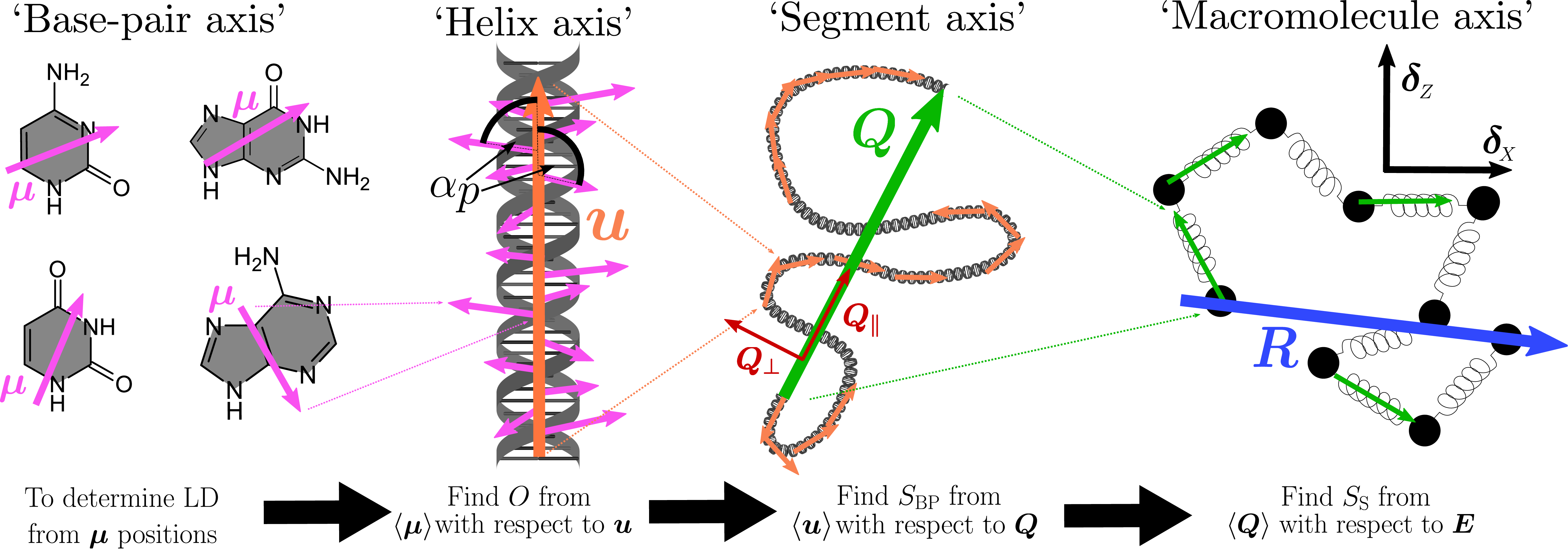}
  \caption[Schematic of different levels at which the orientation parameter $S$ can be calculated]{Schematic of separation of $S$ and $O$ components of $\mathrm{LD}^r$ as per \eref{eq: S and O separation}. Magenta arrow is each transition dipole moment $\bm{\mu}$. The orange arrow is always tangential to the contour of the DNA helix, hence its designation as the helix axis. The green arrow shows the orientation of each spring in our polymer model. Finally, the blue arrow gives the overall orientation of the end-to-end vector, which points between the first and last beads in our model chains (or the first and last monomers in a real polymer chain). The directions of the transition dipole moments $\bm{\mu}$ are purely schematic, and do not represent the real transitions in a DNA helix. Note that for the final schematic, we would have $N_\mathrm{s, BD} = 8$ springs in our coarse-grained polymer model.}
  \label{Fig: LD different levels coarse graining}
\end{figure*}

Consider \fref{Fig: LD different levels coarse graining}, where we have representations of the DNA chain at different length scales.
Beginning at the `base-pair axis' on the left, the overlapping transition moments (with absorption at a particular wavelength of light $\lambda$) are represented by $\bm{\mu}$.
The direction of each of these transition moments can be expressed relative to the orientation of the base pair.
Further, each base pair is embedded within the DNA double-helix, which defines the orientation of a tangent vector $\bm{u}$ at all times parallel to this `helix axis'.
In our polymer models, we cannot individually represent every base pair, but instead split the DNA chain up into $N_s$ segments of length $Q$, each with a vector $\bm{Q}$ pointing along the segment.
Each of these segments is represented by a spring (connecting beads which capture the hydrodynamic friction), and the overall `macromolecular axis' is defined in terms of the end-to-end vector $\bm{R}$.

Furthermore, with our model of the DNA chain with segments $\bm{Q}$, we can write the $\mathrm{LD}^r$ not in terms of the individual transition moments $\bm{\mu}$, but instead in terms of the $\bm{Q}$.
This is possible due to the approximate cylindrical symmetry of the base pairs within the double helix, and the chain contour $\bm{u}$ around each segment $\bm{Q}$ \cite{norden_kubista_kurucsev_1992}.
Specifically, we can separate out the $\mathrm{LD}^r$ into three separate components, which we call the segment orientation $S_\mathrm{s}$, the base-pair orientation $S_\mathrm{BP}$, and the optical factor $O$ \cite{norden2019linear, norden_kubista_kurucsev_1992, Kubista1993}:
\begin{equation}
\label{eq: S and O separation}
    \mathrm{LD}^r = S_s \times S_\mathrm{BP} \times O
\end{equation}
all of which are dimensionless quantities (see also SI for an explicit derivation of a simplified form of this equation).
We now discuss each of these factors in turn.

Firstly, the optical factor $O$ describes the average difference in absorption of light polarised parallel and perpendicular to the DNA helix (i.e. parallel and perpendicular to $\bm{u}$, see \fref{Fig: LD different levels coarse graining}).
This represents an average over both the overlapping transition moments in a base pair at a particular wavelength, and also over all the base pairs along the backbone of the chain.
A commonly used explicit expression for $O$ is given by \cite{norden2019linear}:
\begin{equation}
\label{eq: O formula}
    O(\lambda, E) = \frac{\sum_p A_p(\lambda, E) (3 \cos^2 \alpha_p - 1)}{\sum_p A_p(\lambda, E)}
\end{equation}
where $A_p(\lambda, E)$ is the absorbance of the $p$th transition moment in the DNA chain at some wavelength $\lambda$, and $\alpha_p$ is the angle that transition moment makes with the helix axis $\bm{u}$.
Note that $A_p(\lambda, E)$ (and hence $O$) is a function of the magnitude of the light electric field vector $E$ as per \eref{basic LD equation}, but does not depend on its direction --- the dependence on the direction of $\bm{E}$ (given by the unit vector $\hat{\bm{E}}$) is contained in the $S_\mathrm{s}$ factor.
Other expressions are possible \cite{matsuoka1983linear}, and determination of $O$ for a DNA fragment with either specific, or random base pair sequences has a long history \cite{norden2019linear, norden_kubista_kurucsev_1992, wilson1977dichroic}.
For our purposes, it suffices to assume a random distribution of base pairs, for which $O$ is generally thought to be $O \approx -1.48$ at $260$ nm, the wavelength used in the experiments of Simonson and Kubista with which we will be comparing our simulation results \cite{Kubista1993, norden_kubista_kurucsev_1992}.
This value is derived from assuming the average transition moment is oriented at 86$^\circ$ to the local helix axis \cite{norden_kubista_kurucsev_1992}.

Secondly, the base-pair orientation factor within a segment, $S_\mathrm{BP}$, is derived in a similar way to $O$, only instead of projecting $\bm{\mu}$ onto $\bm{u}$, we project the $\bm{u}$ onto $\bm{Q}$.
Specifically, $S_\mathrm{BP}$ is given by:
\begin{equation}
\label{eq: S_BP}
    S_\mathrm{BP} = \langle (\hat{\bm{u}} \cdot \hat{\bm{Q}}_\parallel)^2 \rangle_Q  - \langle (\hat{\bm{u}} \cdot \hat{\bm{Q}}_\bot)^2 \rangle_Q
\end{equation}
where $\hat{\bm{u}}$ represents a unit vector in the direction of $\bm{u}$, while $\hat{\bm{Q}}_\parallel$ and $\hat{\bm{Q}}_\bot$ represent unit vectors parallel and perpendicular to $\bm{Q}$ respectively.
The ensemble average $\langle \rangle_Q$ is performed for a segment of a particular end-to-end distance $Q$ (not a particular vector $\bm{Q}$) over all possible $\bm{u}$.
For example, the third diagram from the left in \fref{Fig: LD different levels coarse graining} shows a particular chain conformation with length $Q$, but it is easy to imagine that the DNA segment could be bent into a different conformation while maintaining the same $Q$.
To determine $S_\mathrm{BP}$ at this $Q$, we perform a Boltzmann-weighted average \cite{flory1969statistical} over all possible conformations (in our case, this is done numerically using MC simulations).

Crucially, we assume that this average can be performed \textit{at equilibrium} even if our coarse-grained segments are extended due to flow.
This implies that $S_\mathrm{BP} = f( Q )$, where the function $f$ represents the relationship between $S_\mathrm{BP}$ and $Q$ at equilibrium; we then assume that the same functional form $f$ applies to $Q_{\dot{\gamma}}$ at a particular shear rate $\dot{\gamma}$.
This is again equivalent to the assumption that our Boltzmann-average of the possible set of chain conformations in flow at a particular segment extension $Q$ is the same as the Boltzmann-average of the possible set of chain conformations at equilibrium for that $Q$.
This assumption has historically been used in the context of birefringence measurements for polymer chains \cite{kuhn1942relationships, nagai1964photoelastic}.
The function $f$ can be analytically computed for simple chain architectures \cite{wilson1977dichroic, wilson1978flow}, but here we determine it exactly for our model using MC simulations, as described below.
Note that $S_\mathrm{BP}$ must necessarily be equal to $1$ if all the $\bm{u}$ point along $\hat{\bm{Q}}_\parallel$, and equal to $0$ for a random orientation of $\bm{u}$ relative to $\bm{Q}$.

Finally, we come to the segment orientation factor, $S_\mathrm{s}$.
At this point, we project our whole polymer onto the laboratory axes ($\bm{\delta}_X$ and $\bm{\delta}_Z$) along which the dichroism is measured:
\begin{equation}
\label{eq: S_s}
    S_s = \langle (\hat{\bm{Q}} \cdot \bm{\delta}_X)^2 \rangle_{\dot{\gamma}}  - \langle (\hat{\bm{Q}} \cdot \bm{\delta}_Z)^2 \rangle_{\dot{\gamma}}
\end{equation}
for LD with the light polarisation in the laboratory $X$ and $Z$ directions, which are the flow direction $X$ and neutral direction $Z$ (with shear gradient axis $Y$ along which the light is propagated, as per \fref{Couette Cell LD}).
Here the $\langle \rangle_{\dot{\gamma}}$ represents an ensemble average over all segments in the chain, and all chains in our set of simulated chain trajectories, at a particular shear rate $\dot{\gamma}$.
At $\dot{\gamma} = 0$ (a quiescent solution), all the $\bm{Q}$ will point in random directions, and so $S_s = 0$, while a hypothetical flow that perfectly aligns the molecule along $X$ will give $S_s = 1$, which is not necessarily the case as $\dot{\gamma} \rightarrow \infty$ in shear flow.
We calculate $S_\mathrm{s}$ using a BD algorithm as detailed below\deltxt{ and in the SI}.

\subsection{DNA modelling scheme}
\label{BD methods and polymer physics}
\deltxt{
The polymer model and BD simulation scheme has been described in detail in a previous paper \cite{Pincus2023}.
}

\newtxt{
DNA chains have can have contour lengths of tens of microns and relaxation times on the order of seconds.
At these time and length scales, one cannot hope to employ fully-atomistic modelling, and so a coarse-graining procedure must be applied to capture only the relevant physical details necessary to describe the chain conformation in flow.
In developing our model, we follow well-established principles in the coarse-grained modelling of the DNA fragments to ensure accurate determination of the overall chain alignment and stretch under flow \cite{Larson2005review,  Sunthar2005parameterfree, Pincus2023, Saadat2016, sasmal2017parameter, Prakash2019review}.
Our specific DNA model has been described in a previous paper \cite{Pincus2023}, and we will simply give the results here and refer the reader to that work for a detailed discussion.
Our model consists of $N$ beads connected by $N_s = N - 1$ springs, where each spring obeys the so-called FENE-Fraenkel force law \cite{Hsieh2006}, and adjacent springs feel a bending potential which seeks to keep them parallel.
Specifically, if we have a given spring vector $\bm{Q}$ with length $Q$, the force between the connected beads is given by:
\begin{equation}
    \bm{F}^{(c)} = \frac{H(Q-\sigma)}{1-(Q-\sigma)^2/(\delta Q)^2} \frac{\bm{Q}}{Q}
\label{eq:FF_force_eqn dimensional}
\end{equation}
where $H$ is the spring constant, $\sigma$ is the natural length, and $\delta Q$ is the spring extensibility.
Further, we include a bending potential of the form:
\begin{equation}
\label{eqFF: bending potential}
    \phi_{\mathrm{b},\mu}/k_\mathrm{B} T = C (1-\cos{\theta_\mu})
\end{equation}
where $\theta_\mu$ is the angle between adjacent springs, and $C$ is a constant characterising the strength of the potential. 
The real chain has contour length $L_c$ and persistence length $l_p$ as before, so that with $N_s$ segments each segment has a true length $l_s = L_C/N_s$.
To reproduce this chain, we set $\sigma + \delta Q = l_s$ for each spring, then:
\begin{equation}
\label{second H equation}
    H = \frac{k_\mathrm{B} T}{\delta Q^2} \left( \frac{3 l_s^2}{\langle R_\mathrm{DNA}^2 \rangle} - 5 + 5 \frac{l_s}{l_p}\right)
\end{equation}
where:
\begin{equation}
\label{end to end distance WLC}
    \langle R_\mathrm{DNA}^2 \rangle = 2 l_s l_p - 2 l_p^2 \left(1 - e^{-l_s/l_p} \right)
\end{equation}
is the average squared end-to-end distance of the segment of DNA represented by the spring.
Finally, we fit $\sigma$ subject to the above conditions such that $\langle Q^2 \rangle_\mathrm{eq} = \langle R_\mathrm{DNA}^2 \rangle$, which is an implicit algebraic equation which is straightforward to solve numerically.
The condition for the bending potential is in a form originally suggested by Saadat and Khomami \cite{Saadat2016}:
\begin{equation}
    C = \frac{1+p_\mathrm{b,1}(2N_\mathrm{K,s}) + p_\mathrm{b,2}(2N_\mathrm{K,s})^2}{2N_\mathrm{K,s}+p_\mathrm{b,3}(2N_\mathrm{K,s})^2 + p_\mathrm{b,4}(2N_\mathrm{K,s})^3}
\end{equation}
where $p_\mathrm{b,i}$ are specially chosen constants, and $N_{k,s} = L_C/(2 N_s l_p)$ is the number of Kuhn steps per segment $s$.
By including both the bending and spring potentials in this form, we can accurately reproduce the equilibrium end-to-end vector distribution of the real underlying DNA chain given some arbitrary number of springs $N_s$ \cite{Pincus2023, Pincus2022, Saadat2016}.
}

Note that we use the same discretization procedure for both BD and MC simulations, although with different levels of coarse-graining.
Where there is potential confusion, we denote the number of springs in the BD simulations by $N_\mathrm{s, BD}$ and similarly $N_\mathrm{s, MC}$ for Monte-Carlo simulations.
$N_\mathrm{s, BD}$ corresponds to the number of springs at the `macromolecule axis' level as in \fref{Fig: LD different levels coarse graining}, while $N_\mathrm{s, MC}$ corresponds to the number of springs at the `segment axis' level.

\newtxt{
We further include hydrodynamic interactions (HI) and excluded volume (EV) forces between beads.
Hydrodynamic interactions are modelled via the RPY tensor, a regularisation of the Oseen-Burgers tensor, describing how the force on one bead influences the motion of the others:
\begin{equation}
    \bm{\Omega}(\bm{r}) = \frac{3 a}{4\zeta r} \left(A \bm{\delta} + B \frac{\bm{r} \bm{r}}{r^2} \right)
\label{eq:HI tensor}
\end{equation}
where the values of $A$ and $B$ depend on the bead separation:
\begin{subequations}
\begin{equation}
    A = 1 + \frac{2}{3} \left(\frac{a}{r}\right)^2, B = 1 - 2 \left(\frac{a}{r}\right)^2 \text{ for } r\ge 2a
\end{equation}
\begin{equation}
    A = \frac{4}{3}\left(\frac{r}{a}\right) - \frac{3}{8} \left(\frac{r}{a}\right)^2, B = \frac{1}{8} \left(\frac{r}{a}\right)^2 \text{ for } r < 2a
\end{equation}
\label{eq: AandB_RPY}%
\end{subequations}
where $a$ is the effective hydrodynamic bead radius, and $\zeta$ is the bead friction coefficient. 
Note that we usually represent the strength of HI in terms of the parameter $h^*$, essentially a reduced bead radius.
This is given by:
\begin{equation}
    h^* = \sqrt{\frac{k_\mathrm{B} T}{H}} a \sqrt{\pi}
\end{equation}
the form of which comes from its use in the Zimm model with preaveraged HI \cite{bird1987dynamics}.
Excluded volume is accounted for via a Gaussian potential, of the form:
\begin{equation}
\label{Gaussian potential}
    U_\mathrm{Gauss} = \frac{\nu_\mathrm{ev} k_\mathrm{B} T}{(2 \pi d_\mathrm{ev}^2)^{3/2}} \exp\left\{-\frac{1}{2} \frac{Q^2}{d_\mathrm{ev}^2}\right\}
\end{equation}
where $\nu_\mathrm{ev}$ is the strength of the excluded volume potential (with units of volume) and $d_\mathrm{ev}$ is the range of the potential \cite{Ottinger1996, prakash1999viscometric}.
In the limit of $d_\mathrm{ev} \rightarrow 0$, the Guassian potential approaches the delta-function potential.
This form of the EV has the useful feature that the solvent quality, $\tilde{z}$, can be represented exactly in terms of the chain expansion caused by a particular choice of $\nu_\mathrm{ev}$ \cite{prakash1999viscometric}. 
This potential will generally be used in non-dimensional form, with:
\begin{equation}
    z^* = \nu_\mathrm{ev} \left(\frac{ k_\mathrm{B} T}{2 \pi H} \right)^{3/2}
\end{equation}
which allows the solvent quality $z$ to be expressed approximately as:
\begin{equation}
\label{eq: z star}
    z^* = \tilde{z}  \frac{\chi^3}{\sqrt{N}} \frac{4}{3 K(N_{k,s})}
\end{equation}
where $\chi$ is a dimensionless spring length such that $\chi^2 = \langle Q^2 \rangle k_\mathrm{B} T /3 H$ \cite{Sunthar2005parameterfree}, and $K(N_{k,s})$ is a functional correction for semiflexible chains of the form suggested by Yamakawa \cite{Yamakawa2016}:
\begin{equation}
\label{eq: K eqn Yamakawa}
\begin{aligned}
K(L) &=\frac{4}{3}-\frac{2.711}{L^{1 / 2}}+\frac{7}{6 L} & & \text { for } L>6 \\
&=\frac{1}{L^{1 / 2}} \exp \left(-\frac{6.611}{L}+0.9198+0.03516 L\right) & & \text { for } L \leq 6
\end{aligned}
\end{equation}
Therefore, our $z^*$ can be calculated from $N_\mathrm{s, BD}$ and $\tilde{z}$, the latter of which is measured experimentally \cite{pan2018shear}.
In BD simulations, the range of the potential $d_\mathrm{ev}$ can be chosen arbitrarily to maximise computational efficiency, as it does not affect the solvent quality $\tilde{z}$.
Here we have chosen $d_\mathrm{ev}^* = 2 {z^*}^{1/5}$ \cite{Kumar2004Universal, Sunthar2005parameterfree}, where $d_\mathrm{ev}^*$ is the dimensionless range of the potential.
}

\newtxt{
By including all of these physical effects in our equation of motion for the chain, we can derive the following dimensionless Fokker-Planck equation for the evolution of the distribution function $\psi\left(\bm{r}_{1}, \ldots, \bm{r}_{N}\right)$ \cite{Ottinger1996, bird1987dynamics, Prabhakar2004Separation}:
\begin{widetext}
\begin{equation}
\label{eq: Fokker-Planck equation}
    \frac{\partial \psi^{*}}{\partial t^{*}}=-\sum_{\nu=1}^{N} \frac{\partial}{\partial \bm{r}_{\nu}^{*}} \cdot\left\{\bm{\kappa}^{*} \cdot \bm{r}_{\nu}^{*}+\frac{1}{4} \sum_{\mu} \bm{D}_{\nu \mu} \cdot \bm{F}_{\mu}^{\phi *}\right\} \psi^{*}+\frac{1}{4} \sum_{\nu, \mu=1}^{N} \frac{\partial}{\partial \bm{r}_{\nu}^{*}} \cdot \bm{D}_{\nu \mu} \cdot \frac{\partial \psi^{*}}{\partial \bm{r}_{\mu}^{*}}
\end{equation}
\end{widetext}
where $\bm{\kappa}$ is the velocity gradient tensor,  $\bm{F}_{\mu}^{\phi *}$ is the total force on bead $\mu$ due to the sum of the spring, bending and EV forces, and the tensor $\bm{D}_{\nu \mu} = \delta_{\nu \mu} \bm{\delta} + \zeta \bm{\Omega}_{\nu \mu}$ takes into account hydrodynamic interactions between beads $\mu$ and $\nu$.
Quantities have been non-dimensionalized using the following length, time, and force scales:
\begin{equation}
    l_\text{H} \equiv \sqrt{\frac{k_\text{B} T}{H}}, \lambda_\text{H} \equiv \frac{\zeta}{4H}, F_\text{H} \equiv \sqrt{k_\text{B} T H}
\label{Hookean_system}
\end{equation}
}

\newtxt{
For the simulations under flow, we evolve our polymer model in time using a Brownian dynamics algorithm, which solves the nondimensional stochastic differential equation \cite{Prabhakar2004Separation, Pincus2023}:
\begin{equation}
\label{Ito SDE}
    \mathrm{d} \bm{R}=\left[\bm{K} \cdot \bm{R}+\frac{1}{4} \bm{D} \cdot \bm{F}^{\phi}\right] \mathrm{d} t^{*}+\frac{1}{\sqrt{2}} \bm{B} \cdot \mathrm{d} \bm{W}
\end{equation}
where $\bm{R}$ is a $3\times N$ matrix containing bead co-ordinates, $\bm{K}$ is a $3N \times 3N$ block matrix with the diagonal blocks containing $\bm{\kappa}^*$ and others equal to 0, $\bm{F}^{\phi}$ is a $3 \times N$ matrix containing total force vectors on each bead (due to spring, bending, and EV potentials), $\bm{D}$ is a $3N \times 3N$ block matrix where the $\nu \mu$ block contains the $\bm{D}_{\nu \mu}$ tensor components, $\bm{W}$ is a $3\times N$ dimensional Wiener process and $\bm{B}$ is a matrix such that $\bm{D} = \bm{B} \cdot \bm{B}^{\mathrm{T}}$.
The matrix $\bm{B}$ is not calculated directly, but instead the product $\bm{B} \cdot \mathrm{d} \bm{W}$ is evaluated using a Chebyshev approximation, as originally proposed by Fixman \cite{fixman1986construction, Prabhakar2004Separation}.
Additionally, the stochastic differential equation is integrated using a semi-implicit predictor-corrector method with a lookup table for the spring force law, the algorithm for which has been detailed extensively elsewhere \cite{Prabhakar2004Separation, Hsieh2006, somasi2002brownian, Ottinger1996, Hsieh2003}.
}

\deltxt{
We can also include hydrodynamic interactions (HI) and excluded volume (EV) effects in our model, in order to more faithfully capture the interaction of the DNA chain with the solvent at additional computational cost \cite{Prakash2019review,Pincus2023, Ottinger1996, Prabhakar2004SFG}.
}

\newtxt{
To determine the chain conformations at equilibrium, we perform Monte-Carlo simulations without HI or EV.
HI does not affect static properties, and the effects of EV vanish for very short chains of $L \sim l_p$ \cite{Yamakawa2016} (intuitively, a very rigid segment is unlikely to intersect itself, and so the equilibrium distribution is similar between a random and self-avoiding walk).
In the absence of EV, spring orientation and spring length are decoupled, and so it is straightforward to generate Boltzmann-weighted chain conformations according to \eref{eq:FF_force_eqn dimensional} and \eref{eqFF: bending potential}.
Specifically, to generate a chain of $N_s$ segments one simply generates $N_s-1$ random angles from a distribution satisfying \eref{eqFF: bending potential}, and $N_s$ random segment lengths from a distribution satisfying \eref{eq:FF_force_eqn dimensional} (see \cite{Pincus2022} for further details).
}

\deltxt{
A bead-spring chain with a bending potential is used, recovering the contour length $L_c$, persistence length $l_p$ and hence average equilibrium radius of gyration $R_g$ of the real underlying DNA fragment.
This is done for some arbitrary number of segments $N_s$, such that we can specify the level of coarse-graining (where fewer segments is a more coarse-grained chain) in our model.
Fewer model segments for the same physical DNA fragment leads to a more efficient model with respect to the required computational effort to simulate, but a less faithful depiction of the true underlying chain dynamics, despite the bulk equilibrium $L_c$, $l_p$ and $R_g$ being correct.
This level of coarse-graining is represented by the number of Kuhn steps per segment $N_{k,s} = L_C/(2 N_s l_p)$, which intuitively can be understood as how many (twice the number of) persistence lengths of the underlying chain are represented by each spring in our model.
Therefore, we should expect to see our results converge as $N_s$ grows large, or more specifically $N_{k,s}$ becomes small.
We can also include hydrodynamic interactions (HI) and excluded volume (EV) effects in our model, in order to more faithfully capture the interaction of the DNA chain with the solvent at additional computational cost \cite{Prakash2019review,Pincus2023, Ottinger1996, Prabhakar2004SFG}.
}

\begin{figure*}[ht!]
    \centerline{
    \begin{tabular}{c c}
        \includegraphics[width=8.5cm,height=!]{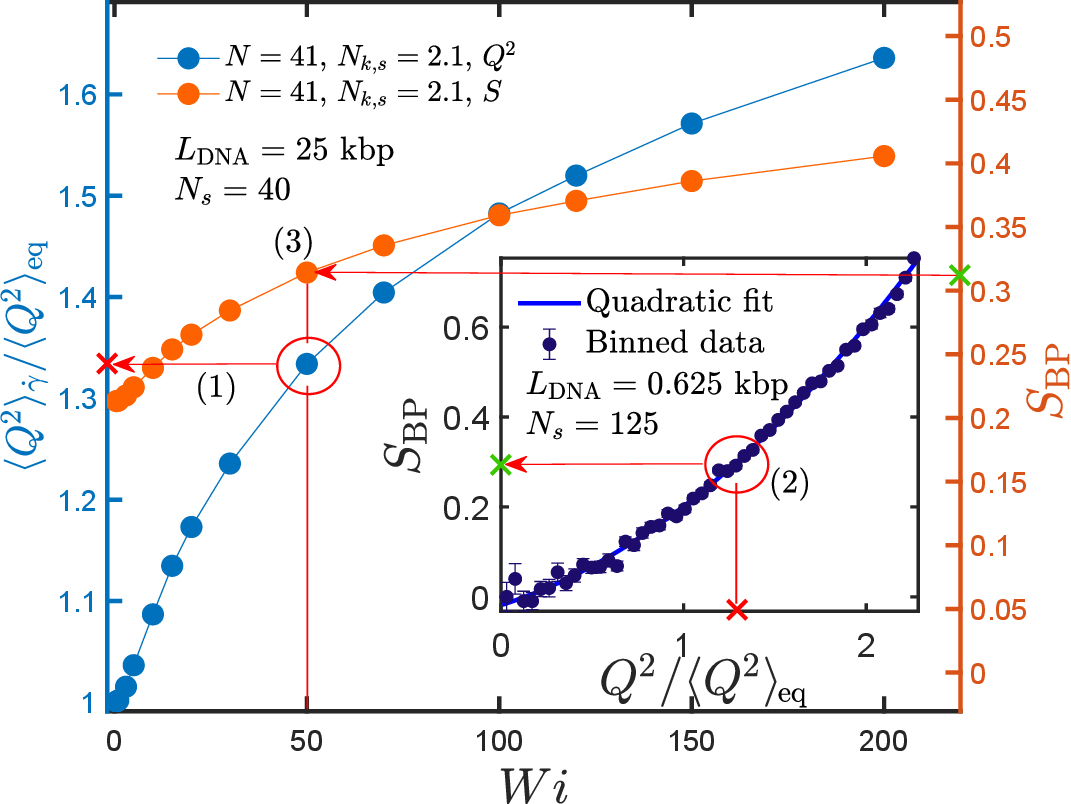} &
        \includegraphics[width=8.5cm,height=!]{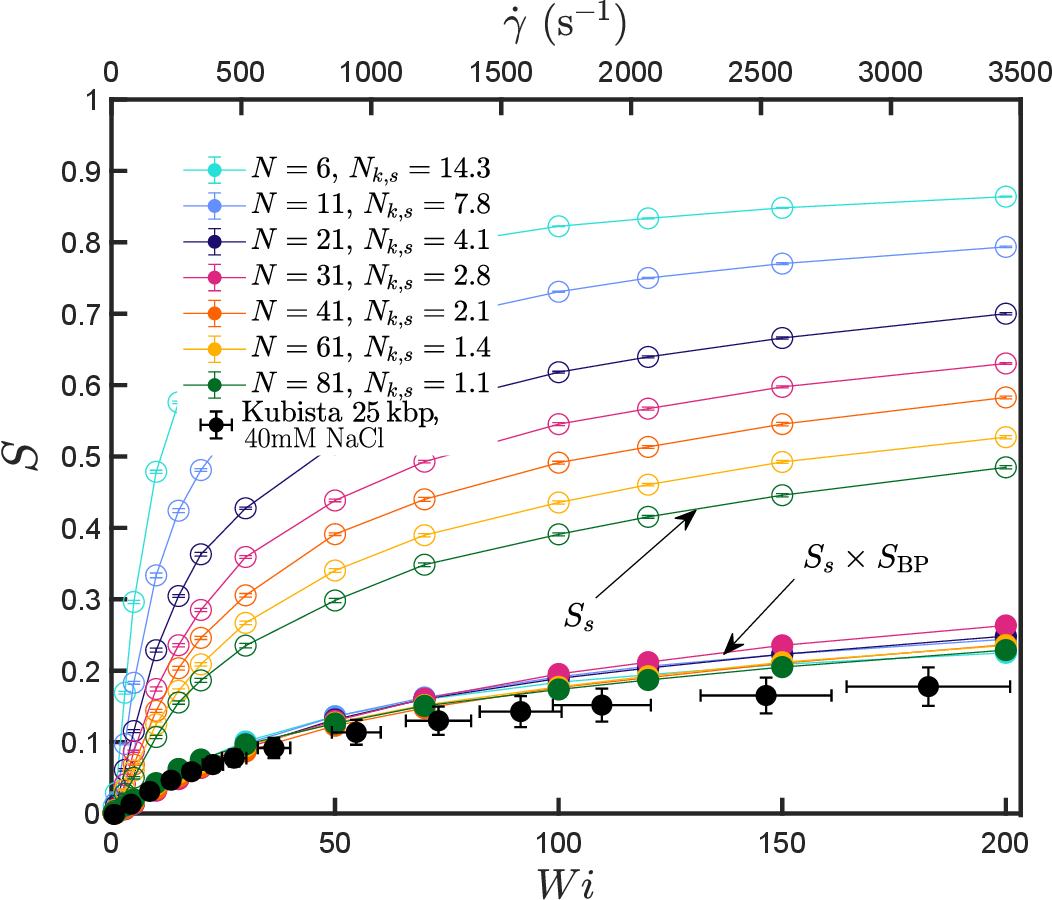} \\
        (a) & (b) \\
        
    \end{tabular}
    }
    \caption[Method of binning data for $S$ against $R^2$ given Monte-Carlo simulations at equilibrium, as well as comparisons with the analytical theory of Wilson and Schellman]{Comparison of Brownian dynamics simulations with experimental data of Simonson and Kubista \cite{Kubista1993}.
    (a)  BD simulation of a 25kbp DNA chain with $N_\mathrm{s, BD} = 40$ springs. Left axis (blue) is $\langle Q^2 \rangle_{\dot{\gamma}} / \langle Q^2 \rangle_\mathrm{eq}$ at each $W\!i$. Inset contains Monte-Carlo simulation at equilibrium for a 625bp segment with $N_\mathrm{s, MC} = 125$, corresponding to a spring length in the BD simulations. To obtain $S_\mathrm{BP}$ (right axis, orange), the steps are (1) calculate $\langle Q^2 \rangle_{\dot{\gamma}} / \langle Q^2 \rangle_\mathrm{eq}$ at a particular shear rate, (2) obtain $S_\mathrm{BP}$ corresponding to that $Q^2$ ratio at equilibrium from the inset, (3) use this to plot $S_\mathrm{BP}$ vs $W\!i$. This sequence of steps is shown explicitly in the figure for $W\!i = 50$.
    (b) BD simulations of $S_\mathrm{s}$ and $S_\mathrm{BP} \times S_s$ for several $W\!i$ of a 25kbp chain following the procedure in the text, alongside the experimental data of Kubista and Simonson \cite{Kubista1993} with $W\!i = \lambda_\eta \dot{\gamma}$, where $\lambda_\eta = 0.058$s. For the experimental data, it is assumed that $S = \mathrm{LD}^r/O$, with $O \approx 1.48$. BD simulations have no HI or EV. \newtxt{Relative} error bars of $10\%$ in $W\!i$ and $15\%$ in $S$ have been added to the experimental data to approximately represent uncertainty in $\lambda_\eta$ and variation in LD due to salt concentration and temperature. Where not visible, simulation error bars are smaller than symbol size.
    }
    \label{fig: bending potential S vs R2}
\end{figure*}

\section{Results}

\subsection{Determining $S_\mathrm{s}$ and $S_\mathrm{BP}$ for DNA under flow}
We now determine $S_\mathrm{s}$ and $S_\mathrm{BP}$ for our DNA chains, focusing first on a 25 kbp chain, beginning with no HI or EV for simplicity, after which we present results for both 25kbp and 48.5kbp chains with HI and EV.
For the 25kbp chain, we choose $N_\mathrm{s,BD} = \{5, 10, 20,30,40,60,80\}$, with spring parameters for each segment calculated as discussed in the \deltxt{SI} \newtxt{methods} with $l_p = 147$ bp \cite{Pan2014ZeroShearVisc} (which given $1$ bp $\equiv 0.34$ nm means $L_c = 8.5$ $\mu$m and $l_p = 50$ nm).
\newtxt{The value of $l_p$ can change with the salt concentration, and we discuss sensitivity of our results to $l_p$ in the SI.
At a constant $\lambda$, errors of a few nm in $l_p$ do not significantly affect the results.}
For each $N_\mathrm{s,BD}$, we calculate the zero-shear viscosity, and hence extract a relaxation time $\lambda_\eta$ \newtxt{(without HI and EV)}.
25kbp DNA has $\lambda_\eta \approx 0.058$s \cite{pan2018shear}, and so the $\approx 10$s$^{-1}$ to $\approx 3000$s$^{-1}$ shear rate range studied by Simonson and Kubista \cite{Kubista1993} corresponds to Weissenberg number $W\!i = \lambda_\eta \dot{\gamma}$ in the range $ 0.6 \rightarrow 175$.
Therefore, determining $S_\mathrm{s}$ is reasonably straightforward - we simulate 500 independent trajectories at a range of shear rates $\dot{\gamma}$ corresponding to $W\!i \approx 0.6 \rightarrow 175$ for each $N_\mathrm{s,BD}$, ensuring that they are run long enough to reach steady state.
Note that this is equivalent to running at $\dot{\gamma} = W\!i/\lambda_\eta (N_\mathrm{s,BD})$, where $\lambda_\eta$ is different for each chain discretisation.
Measurements of $S_\mathrm{s}$ are then taken as per \eref{eq: S_s}, averaged over all segments and trajectories at steady state.

To calculate $S_\mathrm{BP}$ as a function of $Q$, we perform Monte-Carlo simulations.
This is done for each chain discretization $N_\mathrm{s}$ using the same FENE-Fraenkel springs and bending potential as in the BD simulations.
Each spring in our Monte-Carlo simulation has a maximum length of only 5 base pairs (much shorter than $l_p$).
For example, each segment in a $25$kbp DNA chain with $N_\mathrm{s, BD} = 40$ would be represented in the Monte-Carlo simulations by $N_\mathrm{s, MC} = 125$ springs (so if one were to represent the full chain at the level of the MC simulations, we would have 5000 segments of 5 base pairs each, giving $25$kbp).
In order to derive a relationship between $S_\mathrm{BP}$ and $Q^2/\langle Q^2 \rangle_\mathrm{eq}$, we generate thousands of independent chain configurations, then calculate $S_\mathrm{BP}$ for each configuration as per \eref{eq: S_BP}, with $\hat{\bm{Q}}_\parallel$ along the segment end-to-end vector $\bm{Q}$, and $\hat{\bm{Q}}_\bot$ some randomly selected unit vector orthogonal to $\bm{Q}$.
We then bin the data with respect to segment length $Q$ before fitting it to a quadratic as in the inset of \fref{fig: bending potential S vs R2}a. 
At each $W\!i$, $\langle Q^2 \rangle_{\dot{\gamma}}$ is obtained from the BD simulations.
We take $\langle Q^2 \rangle_{\dot{\gamma}} / \langle Q^2 \rangle_\mathrm{eq}$ (step 1 in \fref{fig: bending potential S vs R2}a), assume that it is equivalent to $Q^2/\langle Q^2 \rangle_\mathrm{eq}$ (step 2), and so obtain the required $S_\mathrm{BP}$ with respect to $W\!i$ (step 3).

Finally, we calculate $S$ as the product of $S_s$ and $S_\mathrm{BP}$ as per \eref{eq: S and O separation}, with $S = \mathrm{LD}^r/O$.
The results of this procedure are presented in \fref{fig: bending potential S vs R2}~(b), alongside the experimental data of Kubista and Simonson \cite{Kubista1993}.
It is clear that both the $S_s$ and $S_\mathrm{BP}$ terms are necessary for data collapse for arbitrary $N_\mathrm{s,BD}$, and to qualitatively reproduce the experimental data. 
While $S_s$ depends strongly upon the level of coarse-graining, the product $S_s \times S_\mathrm{BP}$ does not.

\begin{figure*}[t]
    \centerline{
    \begin{tabular}{c c}
        \includegraphics[width=8.5cm,height=!]{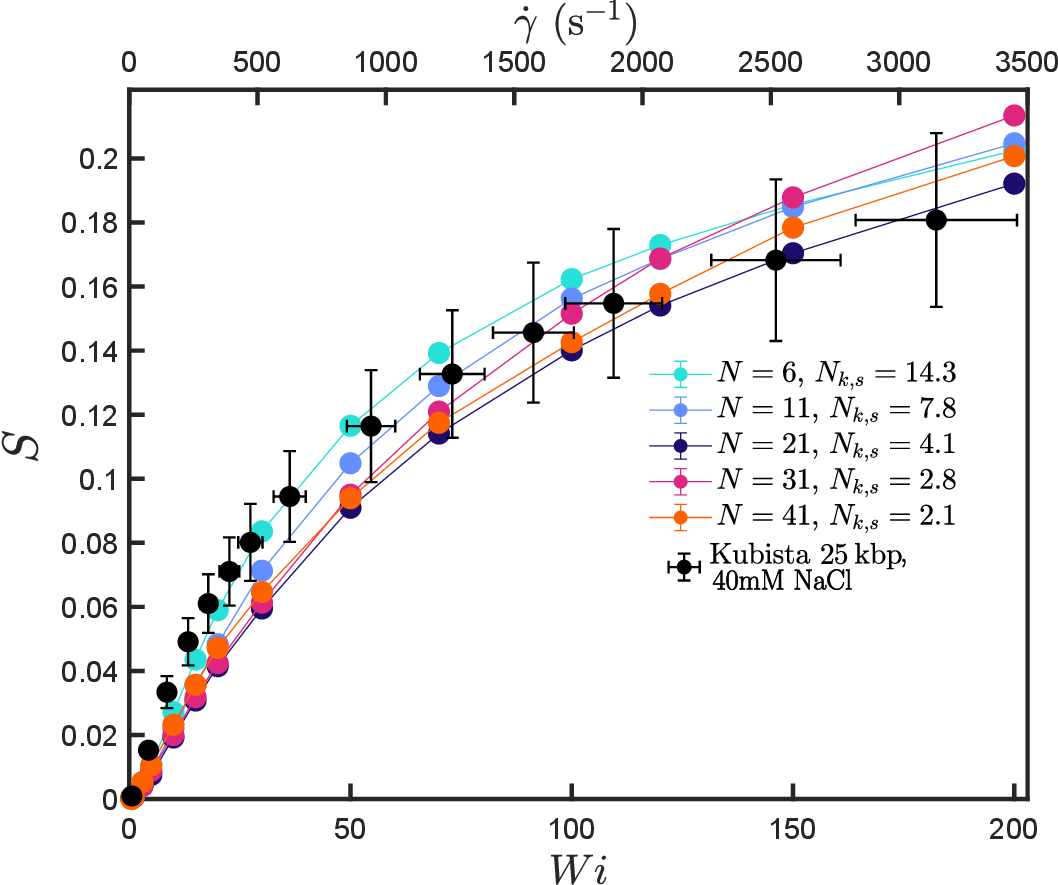}
        & \includegraphics[width=8.5cm,height=!]{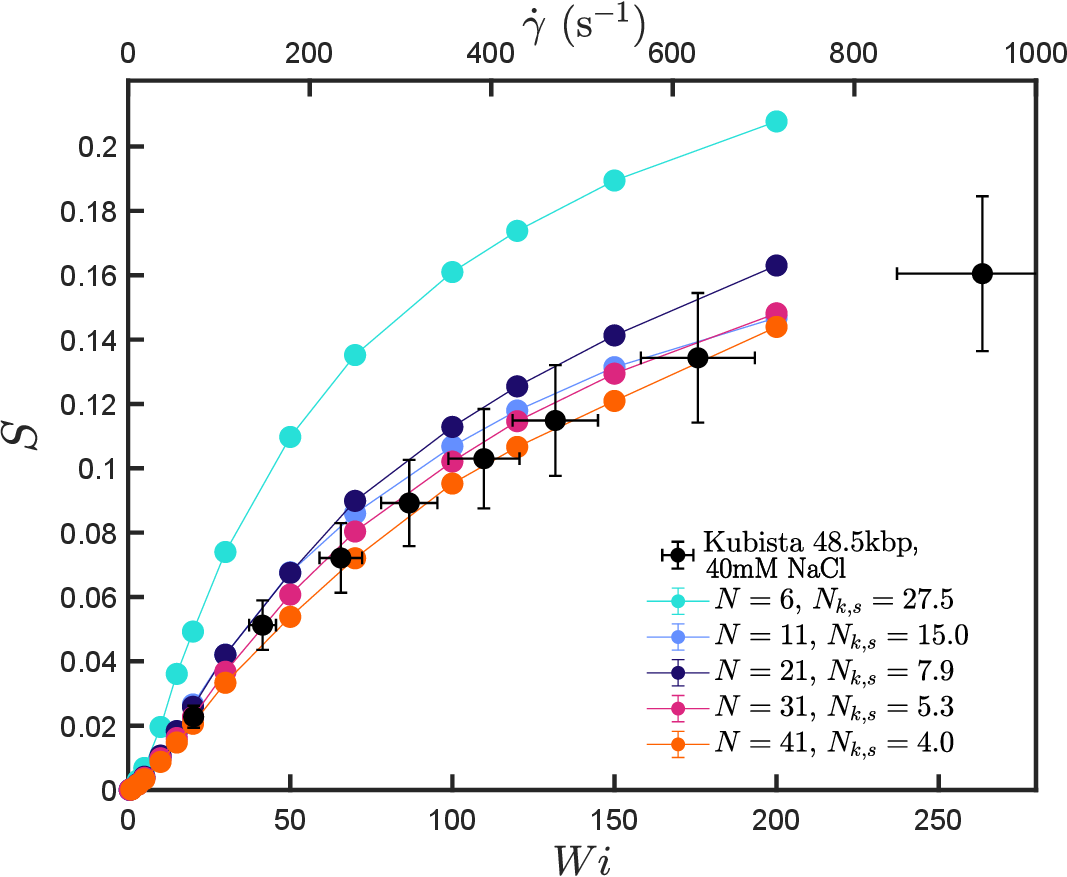} \\
        (a) & (b) \\
    \end{tabular}
    }
    \caption[Comparisons of combined BD-MC simulation predictions with LD data of Simonson and Kubista for 25kbp and 48.5kbp DNA fragments with HI and EV]{Comparisons with the data of Simonson and Kubista \cite{Kubista1993} as in \fref{fig: bending potential S vs R2}~(d). (a) 25kbp DNA with $h^* = 0.3$ and $\tilde{z} = 2$, using $\lambda_{25\mathrm{kbp}} = 0.058$s. (b) 48.5kbp DNA with $h^* = 0.3$ and $\tilde{z} = 2$, using $\lambda_{48.5\mathrm{kbp}} = 0.2$s. For the experimental data, it is assumed that $S = \mathrm{LD}^r/O$, with $O \approx 1.48$. Error bars of $10\%$ in $W\!i$ and $15\%$ in $S$ have been added to the experimental data to approximately represent uncertainty in $\lambda$ and variation in LD due to salt concentration and temperature. Simulation error bars are smaller than symbol size.}
    \label{fig: 25kbp 48.5kbp comparisons}
\end{figure*}

\subsection{Addition of HI and EV}
We can improve our accuracy further by noting that real polymer chains experience hydrodynamic interactions and excluded volume forces between beads.
\deltxt{A brief description of the inclusion of HI and EV in our polymer model is given in the SI, with a more detailed description in previous papers \cite{Prabhakar2004Separation, Pincus2023}}.
The key parameters representing these two microscopic phenomena are $h^*$, the hydrodynamic interaction parameters, and $\tilde{z}$, the solvent quality parameter.
While it is difficult to get an exact measure of the solvent quality for Simonson and Kubista's data \cite{Kubista1993}, Pan et al. have previously found that DNA of similar lengths have values of the solvent quality parameter $\tilde{z}$ in the range $1$ to $3$ \cite{Pan2014ZeroShearVisc}.
Therefore, to investigate the effects of the addition of HI and EV, we have performed simulations with $h^* = 0.3$ and $\tilde{z} = 2$ for both 25kbp and 48.5kbp DNA chains.
While EV will also change the equilibrium distribution for the purposes of calculating $S_\mathrm{BP}$ vs $Q^2$, we have assumed that the Monte-Carlo segments are short enough that the equilibrium distribution is not affected by the presence of EV.
Comparisons with experimental data are given in \fref{fig: 25kbp 48.5kbp comparisons}, showing agreement to within experimental error.
Since simulations with HI and EV require considerably more computational effort (scaling as $N_s^{2.25}$ \cite{Ottinger1996}), we have stopped at $N_\mathrm{s,BD} = 41$, by which point there are diminishing returns to further fine-graining.
\newtxt{
Finally, we note that as for the case without HI or EV, $S_s$ alone does not reproduce the experimental $LD^\mathrm{r}$.
This can be seen in Fig.~S4 in the supporting information for 48.5 kbp DNA.
}

\section{Discussion}

Through systematic coarse-graining in combination with Brownian dynamics and Monte Carlo simulations, we have been able to quantitatively reproduce the LD of DNA in shear flow as shown in \fref{fig: bending potential S vs R2} and \fref{fig: 25kbp 48.5kbp comparisons}.
This is a significant advance which provides a clear demonstration of the progress which has been made in polymer solution modelling since the problem was first studied \cite{wilson1978flow, OdegaardJensen1996}.
Our model requires only five experimentally measured quantities as inputs, namely the DNA contour length $L$, persistence length $l_p$, effective solvent quality $\tilde{z}$, relaxation time $\lambda_\eta$, and optical factor $O$, all of which can be obtained through standard techniques.
Our choices of these parameters in the current work are informed by the best currently available data \cite{Rodger2009, pan2018shear, Pan2014ZeroShearVisc}, but we note that there is some uncertainty regarding the exact value of DNA persistence length as a function of salt concentration, which was found by Simonson and Kubista to significantly affect the measured LD signal \cite{Kubista1993}.
Ideally, one would measure the same DNA solution using several complimentary techniques alongside LD in order to obtain precise values of the aforementioned model inputs, which would provide a powerful further validation of the modelling methodology we have employed in this work. 

\newtxt{
While our method necessarily ignores many chemical details, the coarse-grained approach is justified by the success of the approach, and the length and time scales at play in typical LD experiments.
There are many fully-atomistic simulation techniques which are able to capture the specific chemical details of receptor-ligand interactions, intercalating agents, or the full 3D structure of folded proteins.
However, these codes generally operate over time and length scales many orders of magnitudes smaller than our $\sim 200$ ms relaxation times and $\sim 5$ $\mu$m contour lengths.
As we show through our coarse-grained model in the current paper, there is no need to consider these fine-grained details when simulating a full biopolymer chain.
Further, we consider non-equilibrium behaviour, including the hydrodynamics which are fundamental to polymer behaviour in shear flow.
At these length and time scales, one cannot hope to have a method which fully captures all biological interactions, but instead one must choose what level of detail is appropriate to reproduce and explain the behaviour.
The worm-like chain (WLC) model and polymer physics on which our coarse-graining procedure is built has been successfully applied to a wide range of polymers in the literature \cite{Larson2005review}.
This includes biopolymers such as DNA, RNA, actin filaments, structural proteins, bacteriophages, and more.
}

We \deltxt{finally} wish to briefly discuss why our coarse-graining procedure appears to work so effectively, and the potential limits of its validity.
There is a close analogy between our current methodology and the simplifications for type-A or type-B polymer solutions undergoing dielectric relaxation \cite{watanabe2001dielectric, stockmayer1967dielectric}.
As for type-A or type-B polymers, another way of understanding our procedure is that we are splitting the chain up into many sub-molecules, and then assuming that the end-to-end vector of each submolecule (represented by a single spring in our BD simulations) can be related to the average dipole of that submolecule.
One key assumption made in deriving type-A polymer dielectric relaxation expressions is that results are independent of sub-molecule size, which implicitly depends upon the assumption that the polymer chain obeys equilibrium statistics on length scales larger than the sub-molecules.
It is clear from \fref{fig: 25kbp 48.5kbp comparisons}~(b) that a chain with only $N_\mathrm{s, BD} = 5$ is unable to capture the LD of a 48.5 kbp DNA chain.
Since this corresponds to a chain with $l_\mathrm{s} \approx 10$ kbp, our model is not accurate when segments are coarse-grained to be larger than $\approx 5$ kbp each.
This suggests that at the shear rates investigated in this current work, the flow does not disrupt equilibrium statistics on length scales below 5 kbp, meaning that our key assumptions used in coarse-graining still hold.
This approximation will become less accurate at high shear rates or for large sub-molecules.
This is related to the size of a so-called Pincus blob \cite{pincus1976excluded, Prakash2019review}, where the monomers within a `blob' are internally under equilibrium $\theta$-conditions, with increasing shear rate leading to less monomers per blob \cite{pan2018shear}.
The higher the shear rate, the smaller the submolecule must be in order to still obey equilibrium statistics.
However, for sufficiently small submolecules at sufficiently small shear rates, we should expect that the equilibrium and non-equilibrium partition functions for the submolecules are identical, and so one can assume that $Q$ relates to $S_\mathrm{BP}$ in the same way for both flow and equilibrium.


\newtxt{
Finally, let us address the question of how our current procedure may be useful in quantitative interpretation of LD measurements, specifically in the case of drug binding to biopolymers.
When measuring the LD of a drug-biopolymer system, there are two contributions.
The first is from the biopolymer itself, which we have shown can be decomposed into:
\begin{equation}
    LD^\mathrm{r} = O_\mathrm{biopolymer} \times S_\mathrm{biopolymer}
\end{equation}
where we have further shown $S_\mathrm{biopolymer} = S_s \times S_\mathrm{BP}$ for the current DNA system.
The second  contribution is from the drug molecule.
Small molecules generally do not orient by themselves in shear flow, and so the LD signal will necessarily arise from interaction with the biopolymer backbone.
If this interaction is characterised by some angle which we denote $\alpha$, then we can decompose the LD signal for the drug in a similar way:
\begin{equation}
    LD^\mathrm{r}_\mathrm{drug} = O_\mathrm{drug} \times S_\mathrm{biopolymer} \times f(\alpha)
\end{equation}
where $f(\alpha)$ is some known function of the angle, namely $f(\alpha) = 0.5 (3 \cos^2 \alpha - 1)$ in virtually all cases (since we expect the drug to have an average uniaxial symmetry with respect to the DNA backbone).
Ideally, one would like to measure $\alpha$ using $LD$.
In certain situations, this is possible to do quantitatively, specifically when an internal standard is available such as in certain DNA-ligand complexes \cite{patel2002aryl, rodger1994dna}.
This allows one to determine $S_\mathrm{biopolymer}$ independently during drug binding.
However, in many cases one cannot find $S_\mathrm{biopolymer}$ directly from LD experiments, such that it is only possible to interpret data qualitatively or comparatively \cite{McLachlan2013}.
Our current method outlined in this study has sufficient quantitative accuracy that it can provide a solution to this dilemma.
By determining the $S_\mathrm{biopolymer}$ from independent measurements and BD simulations, one can directly extract quantitative geometric information from LD experiments where a convenient internal standard is not available. 
Future studies will seek to demonstrate this capability directly.
}

\section{Conclusions}
Using a coarse-graining procedure with only DNA contour length $L$, persistence length $l_p$, effective solvent quality $\tilde{z}$, relaxation timescale $\lambda_\eta$, and optical factor $O$ as model inputs, we have been able to quantitatively predict the LD of DNA under shear flow \cite{Kubista1993}.
Inspired by previous theoretical work, our method decomposes the LD signal into three effective length scales, namely at the base-pair level, the coarse-grained segment level, and the overall polymer level. 
These three levels can be separately and independently calculated using a multiscale polymer model to give a more complete description of the polymer dynamics and LD than has been possible with previous approaches.
Although we have compared our predictions to DNA data \cite{Kubista1993}, our method is sufficiently general to be applied to a wide variety of complex macromolecules\newtxt{, many of which have already been studied using Brownian dynamics and coarse-grained polymer modelling \cite{huber2019brownian}.}
Importantly, we have demonstrated that state of the art polymer simulation techniques are a sufficiently mature tool to be used in quantitatively analysing flow dichroism data.
We can now predict how the overall LD signal of DNA should vary upon changes in polymer or solvent properties, for example due to binding of an intercalating dye to DNA \cite{Rodger2009}.
One possibility is to better understand the interaction of potential pharmaceutical drugs with intracellular components such as DNA or structural proteins (actin, spectrin), for which LD is an emerging tool to enable high-throughput drug discovery \cite{Broughton2016}.
In turn, our work demonstrates that LD is a useful tool to validate polymer solution models, hence enhancing our understanding of the rheology of macromolecular solutions.

\begin{acknowledgments}
I. Pincus was supported by an Australian Government Research Training Program (RTP) Scholarship. This research was undertaken with the assistance of resources and services from the National Computational Infrastructure (NCI), which is supported by the Australian Government. This work was also supported by the MASSIVE HPC facility (www.massive.org.au).
\end{acknowledgments}

\bibliography{bibliography}

\end{document}



\title{Supporting information for: \\ 
Flow dichroism of DNA can be quantitatively predicted via coarse-grained molecular simulations}

\author{Isaac Pincus}
\altaffiliation{Present address: Department of Chemical Engineering, Massachusetts Institute of Technology, Cambridge, MA 02139, USA}
\affiliation{Department of Chemical and Biological Engineering, Monash University, Clayton, VIC 3800, Australia
}%

\author{Alison Rodger}
\affiliation{
Research School of Chemistry, Australian National University, ACT 2601, Australia
}%

\author{J. Ravi Prakash}
\email{To whom correspondence should be addressed. E-mail: ravi.jagadeeshan@monash.edu}
\affiliation{Department of Chemical and Biological Engineering, Monash University, Clayton, VIC 3800, Australia
}%

\date{\today}

\maketitle

\deltxt{
Our basic chain model consists of $N$ beads connected by $N_s = N - 1$ springs, where each spring obeys the so-called FENE-Fraenkel force law, and adjacent springs feel a bending potential which seeks to keep them parallel.
Specifically, if we have a given spring vector $\bm{Q}$ with length $Q$, the force between the connected beads is given by:
\begin{equation}
    \bm{F}^{(c)} = \frac{H(Q-\sigma)}{1-(Q-\sigma)^2/(\delta Q)^2} \frac{\bm{Q}}{Q}
\label{eq:FF_force_eqn dimensional}
\end{equation}
where $H$ is the spring constant, $\sigma$ is the natural length, and $\delta Q$ is the spring extensibility.
Further, the bending potential is given by:
\begin{equation}
\label{eqFF: bending potential}
    \phi_{\mathrm{b},\mu}/k_\mathrm{B} T = C (1-\cos{\theta_\mu})
\end{equation}
where $\theta_\mu$ is the angle between adjacent springs, and $C$ is a constant characterising the strength of the potential. 
}

\deltxt{
We have described in detail in a previous paper how one can use these two potentials to model a semiflexible chain with an arbitrary number of springs $N_s$.
Here we will simply give the results, and refer the reader to that work for details \cite{Pincus2023, Pincus2022}.
Assuming that our real chain has contour length $L_c$ and persistence length $l_p$ as before, so that with $N_s$ segments each segment has a true length $l_s = L_C/N_s$.
To reproduce this chain, we set $\sigma + \delta Q = l_s$ for each spring, then:
\begin{equation}
\label{second H equation}
    H = \frac{k_\mathrm{B} T}{\delta Q^2} \left( \frac{3 l_s^2}{\langle R_\mathrm{DNA}^2 \rangle} - 5 + 5 \frac{l_s}{l_p}\right)
\end{equation}
where:
\begin{equation}
\label{end to end distance WLC}
    \langle R_\mathrm{DNA}^2 \rangle = 2 l_s l_p - 2 l_p^2 \left(1 - e^{-l_s/l_p} \right)
\end{equation}
is the average squared end-to-end distance of the segment of DNA represented by the spring.
Finally, we fit $\sigma$ subject to the above conditions such that $\langle Q^2 \rangle_\mathrm{eq} = \langle R_\mathrm{DNA}^2 \rangle$, which is an implicit algebraic equation which is straightforward to solve numerically.
If we add a condition for the bending potential originally suggested by Saadat and Khomami \cite{Saadat2016}:
\begin{equation}
    C = \frac{1+p_\mathrm{b,1}(2N_\mathrm{K,s}) + p_\mathrm{b,2}(2N_\mathrm{K,s})^2}{2N_\mathrm{K,s}+p_\mathrm{b,3}(2N_\mathrm{K,s})^2 + p_\mathrm{b,4}(2N_\mathrm{K,s})^3}
\end{equation}
where $p_\mathrm{b,i}$ are specially chosen constants, we can accurately reproduce the equilibrium end-to-end distribution function of the real underlying DNA chain given some arbitrary number of springs $N_s$ \cite{Pincus2023, Pincus2022}.
}

\deltxt{
\section{Brownian Dynamics and Monte Carlo}
We evolve our polymer model in time using a Brownian dynamics algorithm, which solves the stochastic differential equation \cite{Prabhakar2004Separation, Pincus2023}:
\begin{equation}
\label{eqFF: Ito SDE}
    \mathrm{d} \bm{R}=\left[\bm{K} \cdot \bm{R}+\frac{1}{4} \bm{D} \cdot \bm{F}^{\phi}\right] \mathrm{d} t^{*}+\frac{1}{\sqrt{2}} \bm{B} \cdot \mathrm{d} \bm{W}
\end{equation}
where $\bm{R}$ contains bead coordinates, $\bm{K}$ represents the imposed solvent flow, $\bm{F}^{\phi}$ contains all internal forces on the beads (from springs, bending potentials and excluded volume forces), $\bm{D}$ is a diffusion tensor containing information about bead-solvent friction and hydrodynamic interactions between beads, $\bm{W}$ is a Weiner process representing the stochastic Brownian motion, and $\bm{B}$ is a matrix such that $\bm{D} = \bm{B}\cdot \bm{B}^\mathrm{T}$.
We use a semi-implicit algorithm with $\bm{B} \cdot \mathrm{d} \bm{W}$ calculated indirectly using a Chebyshev approximation, as well as the Rotne-Prager-Yamakawa potential for hydrodynamic interactions \cite{bird1987dynamics}, as detailed elsewhere \cite{Prabhakar2004Separation, Pincus2023, Pincus2022}.
}

\deltxt{
Monte Carlo simulations at equilibrium are performed without HI or EV.
HI does not affect static properties, and the effects of EV vanish for very short chains of $L \sim l_p$ \cite{Yamakawa2016} (intuitively, a very rigid segment is unlikely to intersect itself, and so the equilibrium distribution is similar between a random and self-avoiding walk).
In the absence of EV, spring orientation and spring length are decoupled, and so it is straightforward to generate Boltzmann-weighted chain conformations according to \eref{eq:FF_force_eqn dimensional} and \eref{eqFF: bending potential} (see \cite{Pincus2022} for further details).
}

\deltxt{
\section{HI and EV}
Our strength of hydrodynamic interaction and excluded volume are characterised by the dimensionless parameters $h^*$ and $z^*$ respectively \cite{Pincus2023}.
One can think of $z^*$ as an effective bead-bead repulsive strength, while $h^*$ determines the magnitude of off-diagonal terms in the diffusion tensor $\bm{D}$ in \eref{eqFF: Ito SDE}.
}

\deltxt{
In our simulations, we have set $h^* = 0.3$, but note that the true value may be somewhat lower for short DNA chains \cite{sasmal2017parameter}.
We cannot obtain the EV strength $z^*$ directly from experiments, but instead measure the solvent quality $\tilde{z}$.
In order to match $\tilde{z}$ in our simulations, we can use a so-called soft Gaussian potential for our EV \cite{Ottinger1996, prakash1999viscometric}, and then set \cite{Sunthar2005parameterfree, Pincus2022}:
\begin{equation}
\label{eq: z star}
    z^* = \tilde{z}  \frac{\chi^3}{\sqrt{N}} \frac{4}{3 K(N_{k,s})}
\end{equation}
where $\chi$ is a dimensionless spring length such that $\chi^2 = \langle Q^2 \rangle k_\mathrm{B} T /3 H$ \cite{Sunthar2005parameterfree} , and $K(N_{k,s})$ is a functional correction for semiflexible chain of the form suggested by Yamakawa \cite{Yamakawa2016}:
\begin{equation}
\label{eq: K eqn Yamakawa}
\begin{aligned}
K(L) &=\frac{4}{3}-\frac{2.711}{L^{1 / 2}}+\frac{7}{6 L} & & \text { for } L>6 \\
&=\frac{1}{L^{1 / 2}} \exp \left(-\frac{6.611}{L}+0.9198+0.03516 L\right) & & \text { for } L \leq 6
\end{aligned}
\end{equation}
In this way, we can choose our $z^*$ for some arbitrary $N_\mathrm{s, BD} = N-1$ and $\tilde{z}$, which can be measured experimentally \cite{pan2018shear}.
}

\deltxt{
Note that we have NOT included EV in our Monte-Carlo simulations.
In this case, $N_{k,s}$ is very small since each segment is short and rigid.
If we imagine a real polymer chain with some fixed interaction strength between separated segments (i.e. a fixed $z^*$), \eref{eq: z star} can be inverted to see that $\tilde{z}$ will be vanishingly small for very short polymer segments.
Therefore, the effective solvent quality in our Monte-Carlo simulations is small, and it is not necessary to include EV.
}

\section{Separation of LD optical components}

\begin{figure*}[!ht]
  \centering
  \includegraphics[width=15cm,height=!]{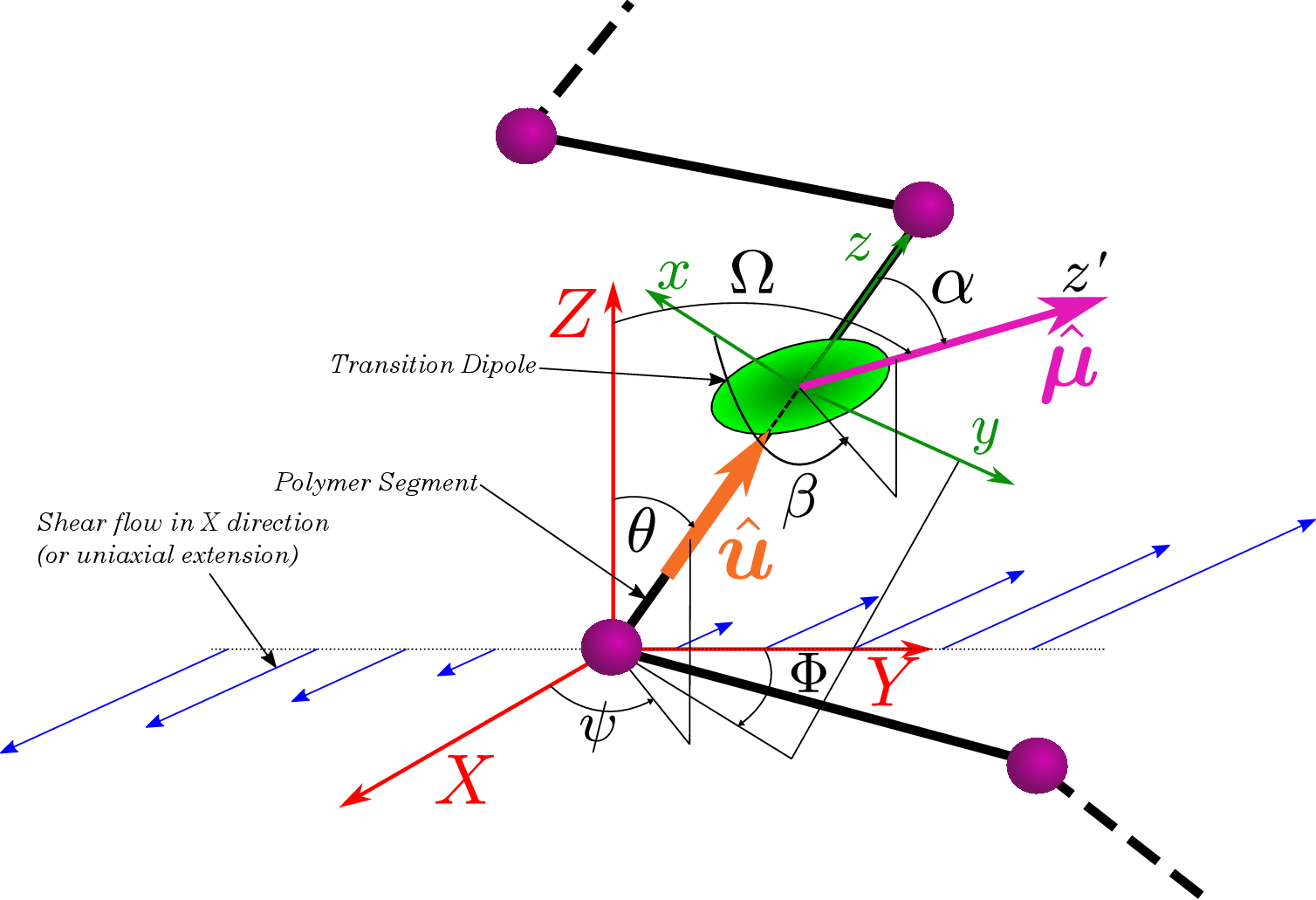}
  \caption[Euler angles for transition dipole of polymer segment]{The transition dipole of a polymer segment is represented as an ellipse, with the long axis oriented at an angle $\alpha$ to $z$. The laboratory frame is defined by the axes $X Y Z$ and is chosen to align with the shear flow direction, while the molecular axis is defined by $x y z$. The transition dipole is aligned along the $z'$ direction. The Couette flow is defined by the shear flow axis ($X$), shear gradient axis ($Y$) and neutral axis ($Z$). Generally, the light will be propagated along the $Y$-axis for LD measurements in a Couette cell, as in Fig.~1 in the main text (meaning that LD is given by $A_Z - A_X$).}
  \label{Polymer Dichroism Euler Angles}
\end{figure*}

Here we wish to show that it is possible to separate the LD signal for DNA into optical ($O$) and orientational ($S$) components.
Specifically, we assumed that we have a polymer chain made up of $N_s$ segments $\bm{u}$ with uniaxial symmetry, each of which have a transition dipole moment $\bm{\mu}$ at angle $\alpha$ to $\bm{u}$.
This is displayed schematically in \fref{Polymer Dichroism Euler Angles}, where the polymer segment orientation is given relative to the background shear flow.
We will not describe explicitly how the $S$ component can be further split into $S_s$ and $S_\mathrm{BP}$ as in the main text, but as we will see, the derivation can be quite naturally continued to arbitrary `levels' of polymer superstructure.

Imagine that we have a coordinate system as in \fref{Polymer Dichroism Euler Angles}, where the unit vector $\hat{\bm{u}}$ of some segment $\bm{u}$ (which points along the molecular $z$ axis) is defined in terms of the elevation $\theta$ and azimuthal angle $\psi$.
$\Omega$ is the angle from the transition dipole moment axis to some laboratory axis along which we measure absorption, in this case the $Z$ axis.
The azimuthal angle is measured from the $X$ axis towards the $Y$ axis.
The transition dipole moment is then embedded at some elevation $\alpha$ and azimuthal angle $\beta$ from the segmental coordinate system $xyz$.

Therefore, the overall unit vector along $\bm{\mu}$, $\bm{\hat{\mu}}$, can be written in terms of the angles $\theta$, $\psi$, $\alpha$ and $\beta$ via four independent rotations.
Essentially, we take a vector in the $Z$ direction, rotate it about the $Y$ axis by $\theta$, then rotate it about the $Z$ axis by $\psi$ - it now points in the direction of $\bm{u}$ (or the axis $z$).
Independently, we can take a vector in the $z$ direction, rotate it about the $y$ axis by $\alpha$, then rotate it about the $z$ axis by $\beta$, so that it now points along $z'$ in the molecular coordinate system.
If we represent these rotations as matrices $\bm{T}_\theta$, $\bm{T}_\psi$, $\bm{T}_\alpha$, and $\bm{T}_\beta$, and further the unit vector along the $Z$ axis as $\bm{\delta}_Z$, then we can define the unit vector $\bm{\hat{\mu}}$ as:
\begin{equation}
\label{eq: rotation matrices}
    \bm{\hat{\mu}} = \bm{T}_\psi \cdot  \bm{T}_\theta \cdot \bm{T}_\beta \cdot \bm{T}_\alpha \cdot \bm{\delta}_Z
\end{equation}
which is the vector:
\begin{equation}
   \bm{\hat{\mu}} = \left(
\begin{array}{c}
 \sin (\alpha ) [\cos (\beta ) \sin (\theta ) \cos (\psi )-\sin (\beta ) \sin
   (\psi )]+\cos (\alpha ) \cos (\theta ) \cos (\psi ) \\
 \sin (\psi ) [\sin (\alpha ) \cos (\beta ) \sin (\theta )+\cos (\alpha ) \cos
   (\theta )]+\sin (\alpha ) \sin (\beta ) \cos (\psi ) \\
 \cos (\alpha ) \sin (\theta )-\sin (\alpha ) \cos (\beta ) \cos (\theta ) \\
\end{array}
\right)
\end{equation}

We first note that $\bm{\hat{\mu}}$ is in fact still a unit vector, which can be verified by calculating its length and applying trigonometric identities - this means that the isotropic absorbance of this system is $1/3$.
In order to find the $\mathrm{LD}_{ZX}$, we simply take the distributional average over the components of $\bm{\hat{\mu}}$
\begin{equation}
    \mathrm{LD}_{ZX} = \langle \hat{\mu}_Z^2 - \hat{\mu}_X^2 \rangle
\end{equation}
where $\langle \quad \rangle$ is essentially an integral over $\theta$, $\psi$, $\alpha$ and $\beta$, which is a linear operation.
In fact, it is possible to explicitly perform this integral for $\beta$, which is uniformly distributed from $0$ to $2 \pi$, and does not depend on any of the other angles.
When we take this integral and normalise the result (which is tedious but straightforward - we have used a mathematica script), we obtain the following expression:
\begin{equation}
   \mathrm{LD}_{ZX} = \frac{1}{16} \left\langle  \left(3 \cos (2 \alpha )+1\right) \left(-2 \sin ^2(\theta ) \cos (2 \psi )+3
   \cos (2 \theta )+1\right) \right\rangle
\end{equation}
Next, note that $\alpha$ is not dependent on the distribution of $\theta$ and $\psi$, so it can be separated into its own term.
By applying the double-angle formula, and simplifying the expression using trigonometric identities, we arrive at the following result:
\begin{align}
\begin{split}
   \mathrm{LD}_{ZX} & = \frac{1}{16} \left\langle  3 \cos (2 \alpha )+1\right\rangle \left\langle-2 \sin ^2(\theta ) \cos (2 \psi )+3
   \cos (2 \theta )+1\right\rangle \\
   & = \frac{1}{16} \left\langle  6 \cos^2 \alpha - 2 \right\rangle \left\langle -4 \sin^2 \theta \cos^2 \psi + 2 \sin^2 \theta + 3 \cos^2 \theta - 3 \sin^2 \theta + 1 \right\rangle \\
   & = \frac{1}{16} \left\langle  6 \cos^2 \alpha - 2 \right\rangle \left\langle -4 \sin^2 \theta \cos^2 \psi + 4 \cos^2 \theta \right\rangle \\
   & = \frac{1}{2} \left\langle  3 \cos^2\alpha-1 \right\rangle \left\langle \cos^2\theta - \cos^2\psi \sin^2 \theta  \right\rangle \\
   & \equiv O \times S
\end{split}
\end{align}
If  we assume that $\alpha$ is a constant $\alpha_\mathrm{eff}$, and further note that $\cos^2\theta \equiv u_Z^2$ and $\cos^2\psi \sin^2 \theta \equiv u_X^2$, we finally obtain an expression in the form $\mathrm{LD}^\mathrm{r} = S \times O$.
A more complicated expression for $O$ in the case of DNA is expressed by Equation~4 in the main text, but the general principle remains the same.
We can also appreciate how this expression could be simplified for a uniaxial extension - in that case, we could eliminate terms by assuming a cylindrical distribution about $X$.

\begin{figure}[!ht]
  \centering
  \includegraphics[width=9.5cm,height=!]{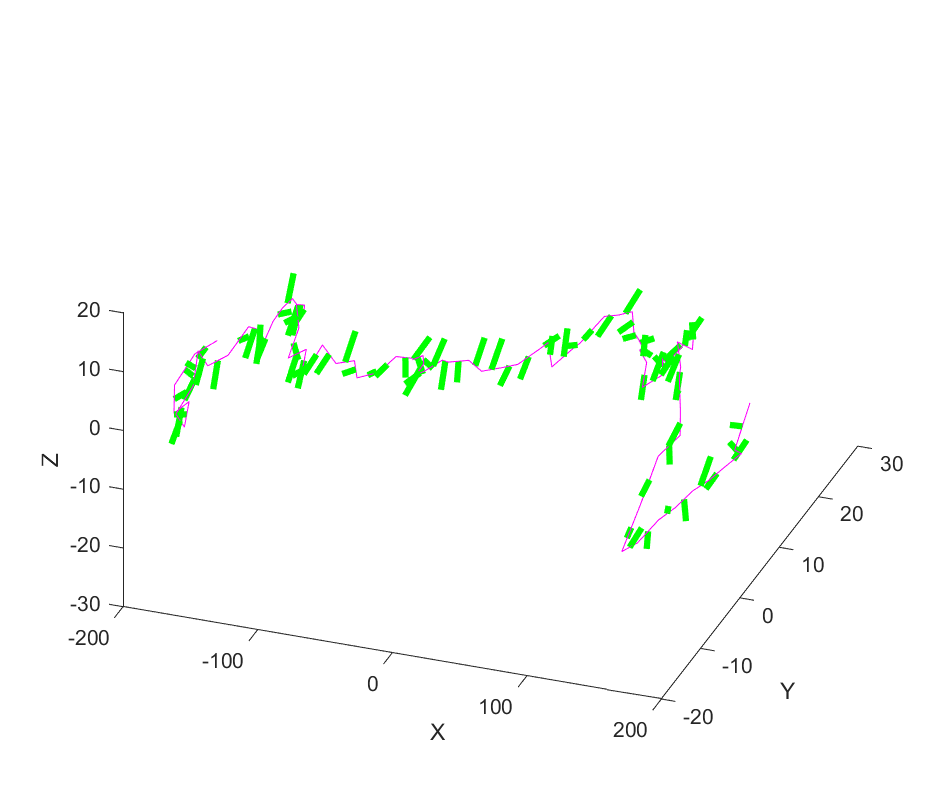}
  \caption[Embedded transition dipole moments in a simulated polymer conformation]{3D conformation of a sample polymer trajectory in magenta. Green lines are the embedded transition dipole moments at an angle of $86^o$ to the segments.}
  \label{Fig: example embedded moments}
\end{figure}

\begin{figure}[!ht]
  \centering
  \includegraphics[width=9.5cm,height=!]{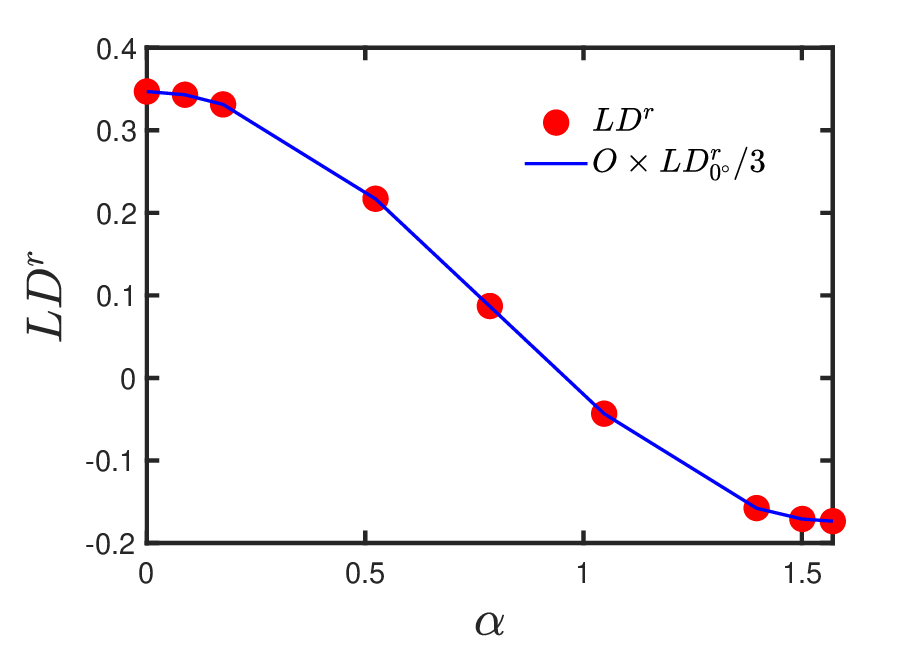}
  \caption[Confirming correct calculation of $S$-parameter in shear flow]{Plot of calculated $\mathrm{LD}^r$ as a function of $\alpha$, both directly calculating the LD (red symbols) and using the $S$-parameter definition for shear flow multiplied by $O$.}
  \label{Fig: alpha plot comparison}
\end{figure}

We can even check the expression numerically for one of our simulations at finite shear rates.
To do so, we embed a transition dipole moment at some fixed angle $\alpha$ and random angle $\beta$ to each polymer segment in shear flow.
This is displayed in \fref{Fig: example embedded moments} for $\alpha = 86^o$, with the transition dipole moments displayed as green arrows.
When one does this for a sufficiently large ensemble of trajectories, and further notes that $S \equiv \mathrm{LD}^r(\alpha = 0^o)/3$, we can iterate over several $\alpha$ and produce a plot such as \fref{Fig: alpha plot comparison}, which confirms that it is sufficient to calculate $S$ and then multiply by $O$ to arrive at the reduced $\mathrm{LD}^r$.
This is of course an analytical certainty given the uniaxial symmetry of our segments and transition dipole moments, but it is a useful check on our procedure for calculation of $S$.

Note that this procedure can quite easily be extended to a further `level', if we have the $\hat{\bm{u}}$ embedded in some more coarse-grained segment $\bm{Q}$.
As long as the distribution of the $\bm{\mu}$ is uniaxial with respect to the $\bm{u}$, and the distribution of $\bm{u}$ is uniaxial with respect to the $\bm{Q}$, we can simply add another transformation matrix to \eref{eq: rotation matrices} corresponding to the transformation from $\bm{u}$ coordinates to $\bm{Q}$ coordinates, then take the corresponding averages.
This gives us the separation in Equation~3 in the main text.

\begin{figure}[t]
  \centering
  \includegraphics[width=9.5cm]{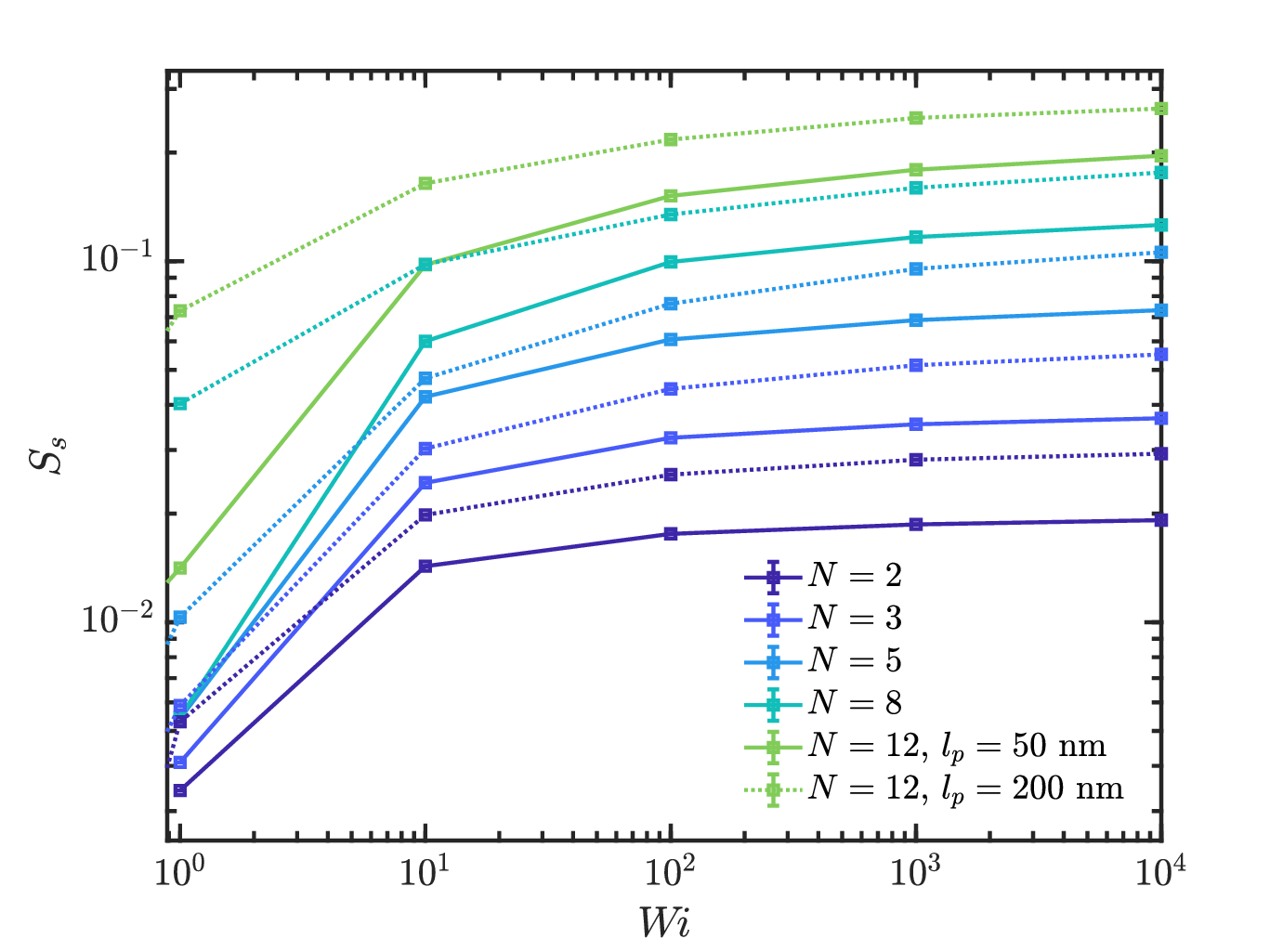}
  \caption{\newtxt{Change in $S_s$ as a function of the inputted persistence length for a hypothetical semiflexible chain with $L_c = 2000$ nm. Full lines represent a chain with $l_p = 50$nm, dotted lines a chain with $l_p = 200$nm.}}
  \label{Example LD changes}
\end{figure}

\newtxt{
\section{Sensitivity to Persistence Lengths}
The rheology and flow orientation of polymers changes significantly with the persistence length.
This is also true for our model, where two polymers with different $l_p$ but the same $L$ will generate different predictions for $S$, both due to the altered spring and bending potentials as well as a change in the relaxation time $\lambda$.
For sufficiently long polymers, the relaxation time should scale with the persistence length to the 3/4 power (which then rescales the effective shear rate $W\!i$).
The specific value of $S$ at a particular $W\!i$ is a more complicated nonlinear function of $l_p$, so it is nearly impossible to make a generic statement about the effects of $l_p$ on $LD^\mathrm{r}$.
In \fref{Example LD changes}, we show some example data for a relatively short semiflexible chain of only 2 $\mu$m, where we see a change in $l_p$ by a factor of 4 consistently leads to a change in $S_s$ of $\approx 30\%$ irrespective of the level of coarse-graining.
Overall, it seems to be the case that small errors, for example, in the experimental value of $l_p$ are highly unlikely to lead to drastic errors in the computed $LD^\mathrm{r}$.
However, differences in polymer architecture or chemistry, for example going from ssDNA to dsDNA, would certainly have a significant impact upon the LD signal computed via our model, since this would affect both the $S$ at a particular $W\!i$, as well as the $W\!i$ itself through the relaxation time $\lambda$ at a particular shear rate.
}

\newtxt{
\section{Separation of components for 48.5 kbp DNA}
}

\begin{figure}[!ht]
  \centering
  \includegraphics[width=12cm,height=!]{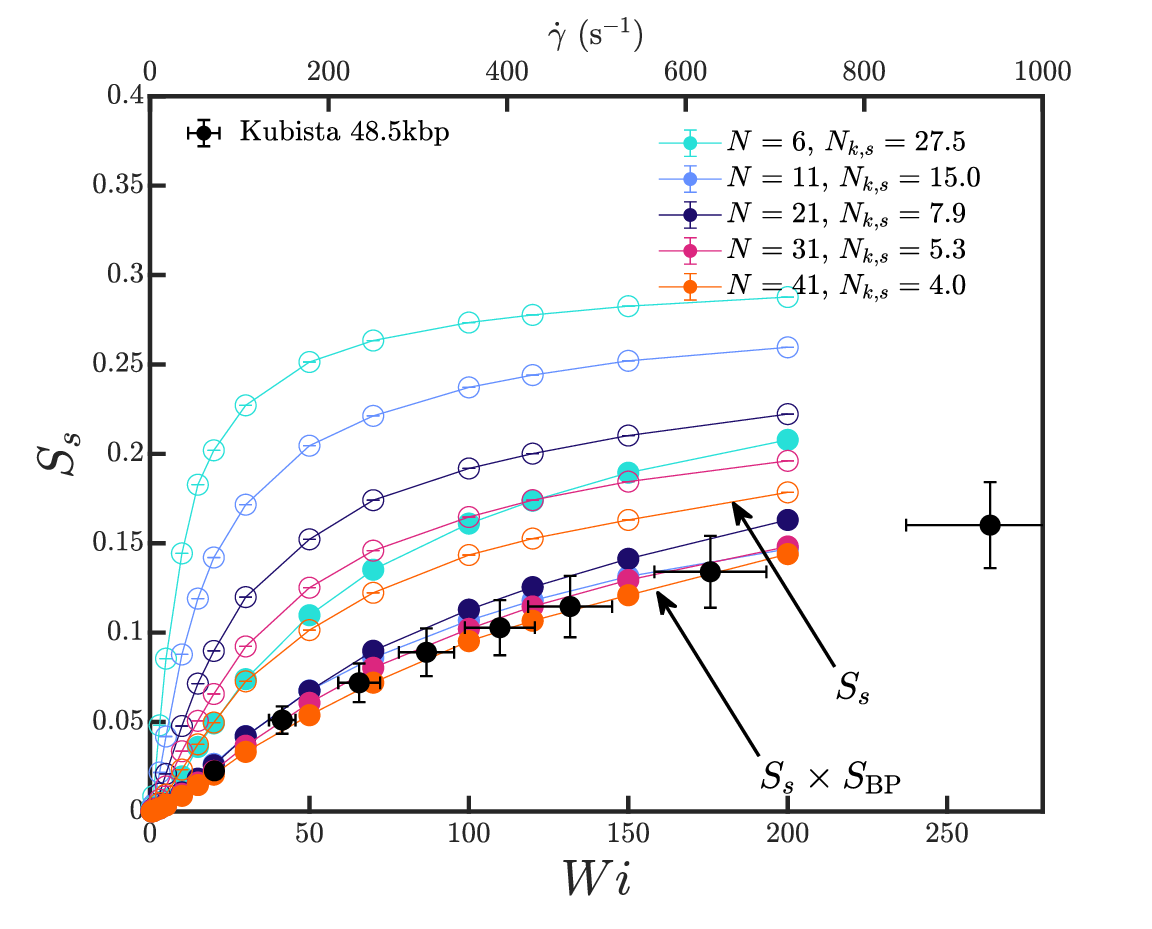}
  \caption[Separation of components]{\newtxt{BD simulations of $S_s$ and $S_\mathrm{BP}$ for several $W\!i$ corresponding to Fig.~4b in the main text. }}
  \label{Fig: 4b separation}
\end{figure}

